# Selection of an engineering institution: Students' perceptions of choice characteristics and suitability under the COVID-19 pandemic

*Corresponding author (*)*
**Prashant Mahajan (*)**
R. C. Patel Institute of Technology, Shirpur.
E-mail: registrar@rcpit.ac.in
ORCiD ID: 0000-0002-5761-5757

**Vaishali Patil**
RCPET's, Institute of Management Research and Development, Shirpur.
E-mail: imrd.director@gmail.com

**Abstract**

Background:
COVID-19 has impacted Indian engineering institutions (EIs) enormously. It has tightened its knot around EIs that forced their previous half-shut shades completely down to prevent the risk of spreading COVID-19. In such a situation, fetching new enrollments on EI campuses is a difficult and challenging task, as students' behavior and family preferences have changed drastically due to mental stress and emotions attached to them. Historically, during crisis situations, institutions have struggled to return to the normal track. Consequently, it becomes a prerequisite to examine the choice characteristics influencing the selection of EI during the COVID-19 pandemic to make it 'normal' for 'new' enrollments.

The purpose of this study is to critically examine choice characteristics that affect students' choice for EI and consequently to explore relationships between institutions' characteristics and the suitability of EI during the COVID-19 pandemic across students' characteristics. Quantitative research, conducted through a self-reported survey composed of a closed-ended structured questionnaire, was purposefully incorporated into the students who recently were enrolled in EIs (batch years 2020-2021) belonging to the North Maharashtra region of India.

Results:
The findings of this study revealed dissimilarities across students' characteristics regarding the suitability of EIs under pandemic conditions. Regression analysis revealed that EI characteristics such as proximity, image and reputation, quality education and curriculum delivery have significantly contributed to suitability under COVID-19. At the micro level, multiple relationships were noted between EI characteristics and the suitability of EI under the pandemic across students' characteristics.

Conclusion:
Bringing 'normality' to 'new' enrollments totally depends on EI's resilience in meeting the needs of diversity in the COVID-19 pandemic situation, which repositions themselves to govern student-centric strategies instituted for the overall suitability of EI under pandemic conditions. The study has successfully demonstrated how choice characteristics can be executed to regulate the 'suitability' of EI under the COVID-19 pandemic for the inclusion of diversity. It is useful for policy makers and academicians to reposition EIs that fetch diversity during the pandemic. This study is the first to provide insights into the performance of choice characteristics and their relationship with the suitability of EIs under a pandemic and can be a yardstick in administering new enrollments.

**Keywords**

Engineering education, choice characteristics, institutional characteristics, students' characteristics, suitability under COVID-19.

**Declarations**

1. *Data availability statement (Availability of supporting data)*

    - The data that support the findings of this study are available from the corresponding author upon reasonable request.
    - All data generated or analyzed during this study are included in this manuscript.

2. *Competing interests*

    - The authors declare that they have no competing interests.

3. *Funding*

    - No source of funding for this research is applicable and hence not reported.

3. *Authors' contributions*

    - All authors have read and approved the final manuscript.

5. *Acknowledgements*

    - Not applicable

# 1. Introduction

Worldwide, engineering education is viewed as a career of upward movement, which has the potential to progressively influence human skill sets (Andreas Blom & Saeki, 2011), social and quality life (Rojewski, 2002), industrial growth, the economy of the country (Cebr, 2016) and the overall development of the country (Downey & Lucena, 2005). Thus, engineering education has proven to be a key factor for the sustainable and profitable development of society. Diversity into engineering study is not just a 'prestige to have' but much more than it. It captivates the changing world by benefitting the complete advantage of a variety of talents and abilities and brings various qualities that add to economic vitality. It encourages global competitiveness through engineering inventions for the benefit of society at large. However, inclusion of diversity and making an upswing for interest itself is a challenge. Reports on declining enrollments (AICTE New Delhi, 2021) in India and diminishing interest and trends (UNESCO, 2010) (Blom & Cheong, 2010) in engineering education worldwide have signaled a warning for the overall development of techno society. In India, the gap between available capacity and actual admissions in undergraduate engineering is widening year by year, leaving 591962 undergraduate seats vacant in 2019-2020. Dashboard of All India Council for Technical Education, New Delhi displayed that approximately 45% of seats remained vacant in the 2019-2020 academic year, which was 38% in 2012-2013. The quality of engineering education in the Indian setting appears to be the lowest compared to that in other developing countries (Loyalka et al., 2014) due to the problems pertaining to awareness, attraction, inclusion of diversity and service quality (Kamokoty et al., 2015; Lakal et al., 2018; Upadhayay and Vrat, 2017). Despite the fact that demand for engineers remains relatively high, few students are willing to pursue engineering education. Notwithstanding the way that demand for engineers remained high, there are few aspirants willing to pursue engineering education.

Selecting an institute, as acknowledged in previous literature, is a subtle and complex process (Regan & DeWitt, 2015a) (Porter & Umbach, 2006) and a mystery process (Freeman, 2012) that involves a vast range of factors (Heathcote et al., 2020), several human capital (Toutkoushian & Paulsen, 2016), and social capital (Hossler et al., 1989a). It affects all that are involved: students, family members, policy makers and institutions of higher education (M. Palmer et al., 2004). It is associated with a multifaceted and inconsistent set of characteristics belonging to students and institutions (Obermeit, 2012) (D. Chapman, 1981). It involves a challenging progression for both (Hemsley-Brown & Oplatka, 2015b) and requires greater efficiency and effectiveness. Decisions regarding 'institute choice' can change students' lives forever (Iloh, 2019) and the performance of institutions. Research on engineering education has not received much consideration and is practically missing on institute selection notions, as the research drift appeared to be inclined towards general higher education addressing psychology, sociology, and economics disciplines (Paulsen, 1990). Today, most engineering institutions (EIs) in India with lower enrollments are in vilest positions due to the absence of students' assessment of their needs. Engineering education is highly contrasted with respect to the multidimensional and changing thoughts of students about the quality of staff and teaching, technology change, course value and delivery, outcome benefits, information and influence effect.

There is certain evidence that higher education (HE) needs to be drastically reformed due to unforeseen situations or crises due to political and economic changes, natural disasters (Schuh & Santos Laanan, 2006) and pandemics (Kim & Niederdeppe, 2013). The literature has shown that during such a situation, higher educational institutions (HEIs) have struggled to return on track. Consequently, it becomes a prerequisite, and there is an emergency need for time to examine the choice characteristics for EIs during the COVID-19 pandemic situation. A survey conducted by The International Association of Universities discovered that COVID-19 will affect enrollment numbers for upcoming new academic years (IAU, 2020), even at local levels. In a country such as India with huge diversity, with many challenges to be faced in the current situation, the 'one-size-fits-all' concept may not be operative. In such a situation, an examination of students' perceptions of choice characteristics holds great practical importance for policy makers and administrators of EIs.

Well informed by the evidence discussed above, the main objective of this study is to critically examine choice characteristics that affect students' choice for EI during the pandemic and consequently to explore relationships between institute characteristics and the suitability of EI during the COVID-19 pandemic across students' backgrounds. The above objective is underpinned by following three research questions referring to the selection of an EI during the COVID-19 pandemic.

1. What are the important characteristics of EIs that attract students to enroll? What are students' perceptions of EI suitability during the COVID-19 pandemic? How they differ by their background?
2. How suitability of EIs is related and governed by traditional characteristics of EIs during the COVID-19 pandemic?
3. During the COVID-19 pandemic, how are students' characteristics and EI characteristics connected in regard to the suitability of EI?

## 2. Literature review

This study embraces a systematic review (Bearman et al., 2012) and progresses gradually through searching, selecting and integrating literature on transparent, structured and comprehensive bases that has drawn an extensive exploration of the origins and evolution of institute selection phenomena. An array of theoretical and conceptual models has been noticed to report the college choice process. The literature review revealed that the college choice process is multifaceted (Hossler et al., 1989) and has reformed over time in accordance with ecological changes (Jackson, 1988). 'Institute choice' takes a culmination position when students' characteristics equalize institutional characteristics, i.e., when the expectation of students meets the offers of institutions and turns out to be a win-win situation after arriving on a common goal (Paulsen, 1990). The integration of students and institutional characteristics produces a collective decision (Huntington-Klein, 2018) grounded on awareness and understanding about institutional facilities, having necessary abilities/performance, and support from family and nonfamily members that affects institute choice decisions (Nora & Cabrera, 1992).

*2.1 Choice characteristics*

The literature review has unlocked several common choice characteristics where institutions' offers interact with students' needs. The power of these choice sets varies and is erratic depending on the economy, politics and socioculture of that country. Hemsley-Brown & Oplatka (2015) discovered that despite ample literature, there is no assured list or index of choice characteristics that confirm students picking up a specific institute and institute, ensuring the inclusion of student groups from diverse backgrounds. Students' characteristics are useful for institutions to include their diversity that suits their culture. On the other hand, institutional characteristics are important for students, as they provide a better match for their expectations and needs. Thus, there is a need to understand both kinds of characteristics as they simultaneously and analogously appear in the 'institute choice' decision-making process. Although ample successful studies such as (Chapman, 1981) (Hossler et al., 1989) (Paulsen, 1990) (Han, 2014) have presented a comprehensive view of 'choice' and have justified adequate attention towards innumerable characteristics influencing selection criteria, engineering studies have been relatively less attended. Nevertheless, multiple studies on other higher educational disciplines have provided some fruitful implications. Institutions to make a pathway for prospective students should understand who are the students and what they expect from them and how their expectations can be met by educational offers (Han, 2014). The following section described at length what is known in regard to choice characteristics owning to students and institutions, which are accountable for institute choice.

## 2.2 Students' characteristics

Students' characteristics are personal measures or factors associated with the students and act as stimuli for selection; however, they are not in the control of the service providers (Hemsley-Brown & Oplatka, 2015a). Numerous researchers have cited the importance of demographic and other personal characteristics in making college choices. The list includes students' demographic characteristics (Kutnick et al., 2018) such as gender; socioeconomic characteristics (St. John et al., 2005) such as social class; geographic characteristics (Kern, 2000) such as place of living; academic characteristics (Hicks & Wood, 2016) such as performance in previous examination; and psychographic and behavioral characteristics (Regan & DeWitt, 2015b) such as interest in terms of priority for majors, delivery, information preferences and human influence.

### 2.2.1 Demographic characteristics

The demographic characteristics of students are observable and hence measurable. The literature has shown that several demographics frolicked a substantial role in determining the likelihood of entry in STEM (Jeffries et al., 2020) (Reay et al., 2001) (Lichtenberger & George-Jackson, 2013a).

As quoted by (Powell et al., 2012), selecting an institute is a gendered process. Gender is found to be the most talked, most researched and most valuable characteristic (Jacobs, 1986) (Turner & Bowen, 1999) (Mansfield & Warwick, 2006) and is referred to as a main predicting factor (Palmer et al., 2004) in the institute choice process. Prior to the 1950s, engineering studies were predominantly regarded as men's circle (Gupta, 2012). There is no single clarification for the varied gender discriminations in relation to institute choice. Females remained persistently underrepresented from engineering studies in most parts of the world due to masculine and feminine traits. Mansfield & Warwick (2006) recognized gender discrepancies in several institutional attributes due to different ecological conditions. Studies segregated the evaluation of gender differences in educational benefits/outcomes (Malgwi et al., 2005a), finance aspects (Cattaneo et al., 2017), influence and human support (Mishkin et al., 2016) (Kelly et al., 2019), cultural aspects (Salami, 2007a), safety and security (Calitz et al., 2020), physical and social atmosphere (Mansfield & Warwick, 2006) (M. Gautam, 2015), socioeconomic context (Gupta, 2012), information sources (Veloutsou et al., 2005b), and behavioral and psychometric settings

(Kolmos et al., 2013) while selecting an institution of their choice. These reflections are analogous in the case of making choices for engineering colleges. The study of Salami (2007) detected human motivation and support such as family (Swan, 2015) as a leading magnetic gadget for female students, while Kolmos et al. (2013) noticed program benefits that tend to inspire male students for their inclusion in engineering study. Presently, in India, female participation is approximately 30% in undergraduate engineering studies (AICTE New Delhi, 2021).

*2.2.2 Socioeconomic characteristics*

Socioeconomic characteristics (SES) signify a mixture of cultural, economic and social backgrounds (McDonough, 1997) (Hicks & Wood, 2016). Socioeconomic status (SES) is the second largest constraint in making institute choice decisions, as evidenced by Hearn (1988) Ball et al. (2002) Thompson (2009). Socially backwards class students belonging to rural areas are less likely to engage in higher education than higher social groups in rural areas. In the case of professional courses such as engineering, the gender effect is stronger in rural regions, and the effect stands for all social groups. The gender gap is more in the socially backward class for their inclusion in engineering studies (Rajasenan, 2014), with a major impact pronounced for females. As reviewed by Jha (2017), families with strong SES supposed to provide better assistance to their children and therefore have shown keen interest in engineering college enrollment (Salami, 2007). Sovansophal (2019) revealed that low-SES families need more human support in the form of career guidance and counseling services for EI selection decisions. Consequently, students' entry to EI for socially, culturally and economically disadvantaged groups is imbalanced and ecologically restricted. This drift is reflected in engineering enrollments, where the percentage appearing in undergraduate engineering studies was relatively very small (17% for academic year 2019-2020) (AICTE New Delhi, 2021).

*2.2.3 Geographical characteristics*

Geographical characteristics vary from one place to another and are contingent on values and beliefs, infrastructure availability, accessibility to information in the place where the institution is situated and the most important contributor in making institute choice (Hardy & Katsinas, 2007). In India, geographic characteristics are determined on the basis of urban and rural populations (Mandelbaum, 1970). Rural populations comprising mostly lower SES are located in villages with low density whereas urban populations consist variety of SES and are found in towns and cities that are highly crowded, but enjoys better facilities on basic human needs, better cultural and greater economy (Hnatkovska & Lahiri, 2013). Rural areas are more likely to be deprived of HE due to geographical isolation, loss of communication with the community (Chakrabarti, 2014), stigma of depression (Nieuwsma et al., 2011) and English language anxiety (Saranraj et al., 2016). Therefore, rural families are deficient in information and awareness about engineering studies. This divergence was also recognized by Matusovich et al. (2020) in an availing engineering study. Therefore, rural students are psychologically dependent on their school's motivation (Young et al., 1997) and sociologically dependent on external support (Chakrabarti, 2014), such as peers, college friends, and college staff, to influence their college choice. The rural population contributes 65.53% of the overall Indian population, and approximately 30% of rural students enjoy careers in engineering.

*2.2.4 Pre-college academic characteristics*

Godwin et al. (2016) (Evans et al., 2020) (Phelps et al., 2018) (Moakler & Kim, 2014) predicted that performance in science and mathematics subjects during the precollege period is important in engineering study. In India, for enrolling in engineering studies, students must score in mathematics and science subjects to get into the merit list. Mathematics has a foremost consideration and weightage in engineering education wherein male students are biologically groomed to outperform female students (Alves et al., 2016).

*2.2.5 Psychological and behavioral characteristics*

These characteristics indicate students' intellectual and behavioral perspectives in regard to their attitudes, personalities, values, learning perceptions, motivation, lifestyles and preferred communication, which are reactive to institute choice decisions (Kotler & Armstrong, 2013). Educational motive is not intrinsic completely (student driving their own behavior) but also refers to the student's extrinsic character (student behavior driven by external stimuli) expressed through psychological and behavioral actions (McInerney, 2013).

Liking for curriculum delivery mode:

(Bernard et al., 2004) (Wladis et al., 2014b) mentioned the importance of the delivery mode through which the curriculum is delivered. Selecting the online mode requires computer self-efficacy and recognition for technology (Lim & Zailani, 2012). Macleod et al. (2017) and Aichouni et al. (2013) mentioned the importance of knowledge delivery through online innovative technology such as ICT. Course delivery in education can be through online mode (no physical interaction), campus mode (in campus face-to-face interaction) and hybrid mode (utilization of both). The knowledge gained during in-campus (on-site) teaching-learning often becomes outdated and is restricted to limited resources and human interference, which needs to be updated and opened up regularly (Sulcic, 2010), particularly for professional studies such as engineering. Online mode has opened new doors for many, as it enables them to learn anytime and anywhere they need at their convenience, without geographical or physical restrictions (Goi & Ng, 2008). Although most of the researchers have conceded online education positively due to its flexible delivery along with (Zia, 2020a), who has witnessed a positive impact of motivation and attitude on the online mode, they found a negative influence of the curriculum and technology due to its inability to demonstrate special technical or mathematical subjects.

Priority for engineering majors:

Economic crises encompassing 'technology and economy' have always influenced the choice of course majors (Chapman, 1981) (Maringe, 2006a) (Ming, 2010). In India, historically heavy-duty industry majors such as mechanical, chemical, civil, and electrical engineering were popular engineering majors; however, with the boom in telecommunication and information technology, course majors linked to electronics, computers and allied courses are gaining popularity. Students' backgrounds and experience are crucial in deciding disciplinary engineering majors (Karataş et al., 2016). Psychologically, major choice is a function of social influence, exposure to new technology, employment information, autonomy to use and drive for technology (Sánchez-Gordón & Colomo-Palacios, 2020). Course majors that guarantee career opportunities, employment returns and academic benefits for students are favorable in choosing institutions (Chapman & Johnson, 1979) (Tas & Ergin, 2012) (Caroni, 2011). Such characteristics are more fascinating for high socioculture and technically skilled students to choose their course majors (Kinnunen et al., 2016).

Computer-related engineering programs are acknowledged universally due to the availability of ample job opportunities, higher salaries and high prestige associated with them (María Cubillo-Pinilla et al., 2006). For this reason, EIs offering computer-related course majors have a better chance of being accepted by prospective students. The choice of engineering course majors depends on gender stereotypes mostly on feminism and muscularity (Bordón et al., 2020). Although modern and emerging majors such as computers and their allied majors are seen as 'female-friendly' (Gupta, 2012), a literature review by (Singh et al., 2007) reported a declining trend of female participation in computer-related majors. Traditional fields, such as mechanical and civil fields, are basically male-dominated majors. An advanced and emerging major (high risk, high returns) is the attraction for students with a high social profile, students located in urban locations and students with high academic backgrounds. On the other hand, traditional majors (low risk in terms of returns) are attractive for socially backward students, students belonging to rural places and low academic performers (Tilak, 2020).

Human influence:

A colony of human influence houses family members as well as non-family members such as relatives, friends, schoolteacher, presently enrolled students, and alumni who directly or indirectly support and motivate institute choice. Prime and firm influence emerges mainly from family (Broekemier & Seshadri, 2000) (Freeman & Brown, 2005) (Salami, 2007a) (Sojkin et al., 2012) (Whiston & Keller, 2004) who have prolonged relations with their financial and emotional support (Godwin et al., 2016b) (Brown et al., 2015) along with precollege schools (Phelps et al., 2018) (Mwangi et al., 2019) and friend influence (Gajić, 2012a) (Wadhwa, 2016a) (Agarwala, 2008) during the initial phase of institute choice. Human influence is a key to lubricating 'moment-of-truth' (Kotler et al., 2009) and cocreating 'word-of-mouth' (Bruce & Edgington, 2008) for prospective students.

Students practiced a truncated flow in engineering studies when family members had limited influence on them. The drive for engineering institute choice is initiated at pre-college i.e. schooling days (Lichtenberger & George-Jackson, 2013b) and its impact is more suitable for females (Martin & Dixon, 2016); however, Gray & Daugherty (2004) noticed that schools are not fully promising for urging students into engineering education.

(AL-Mutairi & Saeid, 2016) (Khanna et al., 2014) while selecting one major among various available engineering majors (Jarvie-Eggart et al., 2020). Students presently studying are the actual teamsters that can fetch prospective students of the same cultural and social experiences they belong to into institute by their motivational interaction (Malgwi et al., 2005a). Word-of-mouth (Bruce & Edgington, 2008) spread by institute students is vital in providing information about location, safety and security, academic standards and teaching quality (Paulsen, 1990a). Alumni are the real human product and experience holders of their institute and are most likely to provide information about institute reputation, campus social life and job opportunities (Singer & Hughey, 2002) (Paulsen, 1990a). Knowledgeable parents are more likely to seek additional information from alumni and existing students about residential arrangement, safety measures when their child is female and quality academic services accessible when their child is a high scorer.

The influence of institute staff is effective (Gajić, 2012b) (Ming, 2010) in preparing prospective students for college campuses, as they are the real performers of curriculum delivery. As revealed by (Paulsen, 1990a), institute staff is valuable in presenting information about institutions' overall characteristics that develop strong relationships with a sense of belonging and confidence in students (Micari & Pazos, 2012).

Information Sources:

Prospective students initially have access to institute controlled information sources such as prospectus, brochure and leaflets (Dawes & Brown, 2002) (Gibbs & Knapp, 2012) (Armstrong & Lumsden, 2000) that actions start-up for their information evaluation (Moogan & Baron, 2003). Students access these sources to obtain first-sight about major options, location and economic aspects, which is useful in differentiating one institute from others (Paulsen, 1990a). To some researchers, these sources are not trustworthy (Bennet, 2006), are misleading (Ivy, 2002), are unethical and insufficient (Chapman, 1981) and now a days trailing their audiences. Based on these sources, potential students cannot decide to enrol in particular engineering institutions.

Potential students can resolve their doubts on the spot by face-to-face communication: on campus or off-campus by obtaining additional knowledge and interacting with institute staff. Institutes arrange face-to-face interaction via campus visits, educational exhibitions, precollege visits and conducting seminars at prominent places. Potential students receive information from institute visits to their schools or coaching classes, institute seminars (Yamamoto, 2006b) (Gajić et al., 2017) and educational exhibitions (Dawes & Brown, 2002) (Bodycott, 2009) (Goff et al., 2004) that are promising for them in creating initial interest in the institute.

Potential students admitted the institute website (Berge et al., 2019) (Kotler et al., 2002) (Yamamoto, 2006b) (Simões & Soares, 2010) (S. Briggs & Wilson, 2007a) to effectively shape their choice decisions. They found it easy to read and understand with in-depth content (Tarafdar & Zhang, 2005) that demonstrates several detailed information about current happenings, campus life and professional roles; however, students cannot make immediate communication with the institute.

Due to its rapid and detailed delivery of information, social media is an effective promotional tool (Chauhan & Pillai, 2013) (Turner, 2017); therefore, potential students are comfortable using it for acquiring their educational goals (Ferry et al., 2000). Social media is effective in making two-way interactive communication that engages prospective students, current students, alumni, and other stakeholders for developing prolonged relationships, continuous engagement (Chugh & Ruhi, 2018) that houses trust and relationships (Khanna et al., 2014) with no constraints on time and cost (Kowalik, 2011). Murphy & Salomone (2013) views social media as a useful tool for collaborative and creative work and a facilitator to sustain interactions, while Rutter et al. (2016) cites it as an important branding tool that is successful in the co-creation of 'word-of-mouth' that directs students' pathways towards that institute.

## *2.3 Institute's characteristics*

Institutional characteristics are the selection criteria that appeal to prospective students to enroll themselves by making evaluations. These characteristics are clustered on financial vs non-financial offers, academic vs non-academic facilities and services and tangible vs intangible factors (Hossler et al., 1989b) (Yamamoto, 2006a), which are reviewed as under.

*2.3.1 Proximity*

Proximity relates to nearness to the distance of college from home. Being close to college is a significant factor in college choice (Griffith & Rothstein, 2009) (Turley, 2009), particularly evidenced for females (Sanders, 2005) and low socioeconomic groups (Frenette, 2006), and increases the chance of students' college going (López Turley, 2009). Distance travel is associated with cost, time and efforts (Chapman, 1981). The proximity in regard to engineering college accompanies that such savings can provide extended homework hours and enough time for social involvement.

*2.3.2 Location and locality*

Location and locality is a structure of ambient conditions, spatial layout and functionality (Bitner, 1992) that is a swaying character in making institute choice (Mahajan et al., 2014) (Weisser, 2020) (Gibbs & Knapp, 2012). Location labels as the site of the institute and its connectivity from home, while locality refers to culture, amenities and facilities available in surrounding place wherein the institute is located. It is credited with suitability, vicinity, attractiveness, accessibility, cost-effectiveness, safety and security (Hannagan, 1992; Kotler & Fox, 1995). A location with safe and secured amenities for females (Chanana, 2000) and a cost-effective locality for economically weaker students are apprehension issues in institute choice. In engineering education, physical and digital locations (Ho & Law, 2020) are important concerns for the involvement and interactions of all stakeholders.

*2.3.3 Image and reputation*

The image and reputation of an institute in public minds plays a significant role in differentiating colleges (Imenda et al., 2004) and is measured as one of the topmost characteristics affecting institute choice (S. Briggs, 2006) (Lafuente-Ruiz-de-Sabando et al., 2017) (Wadhwa, 2016b). However, it is a complex spectrum of small reputes allied with quality dimensions though academic or nonacademic. Institute rankings, brand and academic quality and word of mouth are interconnected to contribute to image and reputation (Lafuente-Ruiz-de-Sabando et al., 2018). In the review of the literature (Hemsley-Brown & Oplatka, 2015c) and in most of the research (S. Briggs, 2006) (Gajić et al., 2017) (Maringe, 2006b), the image and reputation of institutes provide the first impression that embosses the minds of decision makers, even if nobody has a confront with an institute.

*2.3.4 Faculty profile*

Faculty profiles in terms of their qualification, skills, competency and experience (Imenda et al., 2004) exert a significant impact on students (Soutar & Turner, 2002) (Lau, 2016) (S. Briggs, 2006) (Mazzarol & Soutar, 2002). (Magnell et al., 2017) mentioned that faculty experience, qualification and attitude are crucial in developing extensive links with industry and that academic professionals are crucial for the strong delivery of engineering curricula. Faculty ought to be profiled with high-quality teaching (Woolnough, 1994) and well designers (De Courcy, 1987). Similarly, they should be well-inspired, well informed, passionate, open minded, and responsive (Voss et al., 2007) to transform knowledge and to assist students in real-world exposure (Bhattacharya, 2004).

*2.3.5 Alumni image*

Alumni are the tangible products of educational services; hence, alumni concerns are an important criterion in measuring engineering college performance considered by accreditation bodies in India. Alumni achievements are often exploited to exemplify the importance, eminence and brand image of a college (Saunders-Smits & de Graaff, 2012) and criteria for selecting a college (Abdullah & Saeid, 2016) (Ho & Hung, 2008). A network of 'old boys' is always utilized to attract prospective students as a role model because of the impact they have had on society. Notable alumni represent guaranteed resources for brand identification (Palmer et al., 2016) for colleges. Historically, alumni images with economic, market and social standing have at all times added glory to the reputation of their college and hence befit benchmarking standards for prospective students (Pucciarelli & Kaplan, 2016).

*2.3.6 Campus placements*

Job opportunities or employment prospects are the potential outcomes and benefits that prospective students and their families seek (Khanna et al., 2014) (Maringe, 2006a) from investing their time, efforts and money in higher education (Hemsley-Brown & Oplatka, 2015c). The transition from education to employment is the straightforward motive of every student opting engineering study (Baytiyeh & Naja, 2012) and has been verified to be one of the most influencing characteristics in choice decisions (Soutar & Turner, 2002), specifically for male students (Malgwi et al., 2005b). Most premium engineering institutions have separate placement offices to deal with students' employment and upholding alliances between industry and academia. Campus placement strategies should boost employability skills (Markes, 2006) and accelerate industry-academia connections (Baytiyeh & Naja, 2012), which are preliminary requisites for creating employment opportunities for engineering students.

*2.3.7 Quality Education*

Quality of education is a prime, discriminating, and prominent characteristic of an engineering institute consigned to stay ahead in competition and to make a place in the minds of stakeholders. Several studies (Pandi et al., 2014) (Upadhayay & Vrat, 2017) (Sakthivel & Raju, 2006a) (Jain et al., 2013) have emphasized the importance of the quality aspect, which is responsible for the qualitative and holistic development of engineering students and important in making institute choice decisions (Paulsen, 1990c) (Kallio, 1995) (Mourad, 2011) (McCarthy et al., 2012). From India's perspective, several items, such as academic standards, industry linkages, campus placements, research and consultancy, accreditation and financial standing, contribute to the quality of education, as mentioned by (Mahajan et al., 2014). Furthermore, for some researchers, it refers to the quality of the curriculum (Izquierdo, 1993), course delivery (Trum, 1992), infrastructure facilities (Sayeda et al., 2010), faculty and staff (Gambhir et al., 2013), quality services (Viswanadhan, 2009), and academic and nonacademic concerns (Jain et al., 2013) (Owlia & Aspinwall, 1998). Overall, it has a two-fold effect in terms of tangible and intangible outcomes (Natarajan, 2009) that accounts for the quality of education for engineering institutions. The presence of a high-scoring group of students (Horstschräer, 2012), faculty profile (Loyalka et al., 2014), college reputation (Abdullah, 2006), and word-of-mouth (Sun & Qu, 2011) are the talked indicators of quality in engineering education.

*2.3.8 Infrastructural and facilities*

The importance of infrastructure and facilities is mentioned in numerous studies, such as (Nyaribo et al., 2012) (Sahu et al., 2013) (I. F. Price et al., 2003). It consists of college buildings, equipment, IT infrastructure and amenities, which are tangible possessions reflecting the capacity of institutions that streamline the performance of curriculum delivery (Palmer, 2003). It can provide love-at-first-sight for institutions and on-the-spot evidence for prospective students (Philip Kotler et al., 2002) that assist selection at first sight. Delivering a curriculum without the existence of infrastructural assets and facilities is not possible for engineering institutes, as delivery is more technical in nature. More are the facilities, more will be the enrollments.

*2.3.9 Safe and secured environment*

Safety on the college campus means the provisions made to ensure the wellbeing of students in regard to their residential, physical health, and life concerns (Ai et al., 2018), whereas security, as a broad term, covers physical, social, and economic dimensions by protecting human rights, emotions and cultural values (Calitz et al., 2020). Studies such as (Soutar & Turner, 2002) (Elliott & Healy, 2001) (Peters, 2018) have exposed that students contemplate it based on wellbeing and humanize culture, principles, tradition, and socioeconomic characteristics allied with institute choice decisions (Calitz et al., 2020). The students feel comfortable with the health services, sanitization services, fire safety and emergency provisions delivered by the institute (Sakthivel & Raju, 2006b). The literature shows that it was the most essential character in making institute selection but now has observed a mandatory obligation to mentally and physically support female, minority, rural and backward class students.

*2.3.10 Curriculum delivery*

In engineering education, curriculum delivery is the most important characteristic of institutions (Gajić, 2012); however, it has become a matter of debate worldwide due to its lack of ability to attract diverse students, promote professionalism (Berjano et al., 2013) and adopt a realistic approach (Mitchell et al., 2019). It executes a planned practice of teaching and learning influenced by intangible services with tangible facilities that ensures consistent provision to transfer and monitor disciplinary knowledge to meet the needs of diverse students (Case et al., 2016). Engineering institutions can bring glory to the institute if delivered as per the needs but can take vilest situations if not handled properly. Curriculum delivery is often bordered by mixed outlines such as online (Alawamleh et al., 2020), hybrid (Tan, 2020) (Sia & Adamu, 2020) and onsite delivery. Although all have their own advantages and disadvantages in regard to theoretical, practical, technology, human involvement and skills approaches, the degree to which it processes knowledge via gaining, accessing, practicing and implementing is more important (Shay, 2014).

However, curriculum delivery was found to be the first priority in selecting an engineering institute in most studies, such as (Moogan & Baron, 2003). Mostly, the course structure of engineering is framed and designed by affiliating universities, and colleges have to deliver it with value addition and enrichment for effective transformation (Walkington, 2002). To attract diversity (Hemmo & Love, 2008), the delivery of engineering curricula is considered the backbone of transforming engineering knowledge into practical applications.

*2.3.11 Value for money*

Cost is a tangible characteristic and is deemed to be financial anxiety for students and affects college choice decisions, hence enrollment and college income. In engineering studies, the nature of costs is differential and includes tuition, travel, residential and food costs, and day-to-day academic costs, which are more expensive than other higher education methods. Naturally, as students rely on their family to invest family income in their education, it may imitate a negative impact on engineering enrollments (Stange, 2015). Some studies have tried to exhibit the cost of education as a package of rewarding benefits entailing value and quality (Jonathan Ivy, 2008) (Joseph et al., 2005), time and effort (Kotler & Fox, 1985), effort and opportunity (Wu et al., 2020).

Some hands of work have documented the cost of education as immaterial in institute selection when services are concerned with quality (Yoon, 2019), associated with high earning benefit (S. Briggs & Wilson, 2007a) and value for money or benefit sought (Foskett & Hemsley-Brown, 2001) (Maringe & Gibbs, 2009) (Kotler & Fox, 1995) (Holdsworth & Nind, 2006) (DesJardins & Toutkoushian, 2005).

## *2.4 Change: Engineering education under the COVID-19 pandemic situation*

### *2.4.1 General impact*

COVID-19 is the disease triggered by coronaviruses in December 2019 that causes respiratory illness spread though small saliva in the form of droplets and aerosols arising out due to close human contacts (Sohrabi et al., 2020) (Ciotti et al., 2020). Very little literature has come up within the reach of accessibility and suitability of higher education (HE) during the pandemic. The pandemic has stressed educators to provide better safety, security and healthcare solutions that regain existing and prospective students. It has taken out higher education by storm and hence turns out to be the most challenging condition in the history of engineering education. As indicated by the World Health Organization, physical and social distancing is the only credible way to constrain its spread.

### *2.4.2 Curriculum changeover*

In the past, during situational crises, studies such as (Rosenthal et al., 2014) discovered curriculum delivery, and (Kim & Niederdeppe, 2013) suggested students' support system as an important factor in normalizing the situation and continuing pedagogy. During different natural crises that appeared earlier in the history of HE, online education was proven to be effective (Rosenthal et al., 2014). As a result, ICT-based online delivery is increasingly becoming an essential requirement for online education (Jayalath et al., 2020) that promises to be more elastic towards pandemic safety measures. With travel restrictions, pandemics will hamper the mobility of students from one place (native place) to another place (college place) (Xiong et al., 2020), which has made educational policy makers stop 'onsite' physical activities and move 'online' (Tesar, 2020). The online mode is conceivably seen as the 'continuing education further' (Settersten et al., 2020) (Gautam & Gautam, 2020; Naziya Hasan & Khan, 2020; Liguori & Winkler, 2020). (Toquero, 2020) has expressed an emergency need of strengthening practices beyond traditional classrooms and reinforcing curricula in accordance with the learning needs of students. However, after the start of 'online' momentum, it has been simultaneously condemned adversely (K. Bird et al., 2020) (Zia, 2020b) (Rosenberg et al., 2020) (Najmul Hasan & Bao, 2020) for various reasons. Few researchers have talked about hybrid/blended delivery (Rashid & Yadav, 2020; Sia & Adamu, 2020) as a primary solution to continuing pedagogy, but without any analytical base.

*2.4.3 Students' changeover*

The US-based research of (Aucejo et al., 2020) has shown that the effect of the pandemic is extremely heterogeneous and more significant for SES groups. However, a recent study by (Mae, 2020) signalled positive signs for US families on HE plans that have not been affected but are ready to invest in HE as planned earlier to pandemic.

(Aristovnik et al., 2020) revealed that the pandemic with emotional as well as personal life has also affected certain demographic characteristics and SES, such as academic work/life and financial issues. The author further discloses that the pandemic has brought out changes in the habits of students and has drastically changed the role of college management in implementing new emergency policies that were never there previously. Overall, recent studies have cited that the pandemic has reordered the self-beliefs and perceptions of students as well as their priority for college offers. This change is realized more for an education that will take place at a far distance (Mok et al., 2021).

*2.4.4 Institute changeover*

(UNESCO, 2020) judges that in times of crisis, no one can assume an 'online' mode to deliver the entire curriculum as usually provided by the onsite mode; therefore, the delivery has to be redefined or reduced and replaced or enhanced. On the other hand, students must be engaged particularly to avoid academic, social and emotional loss of students, particularly those who are female, resident in rural areas and belong to SES families. (WHO, 2020) expressed an urgent need for policy reforms and multilevel coordination that sustain the mental health and social emotions of students. (Chadha et al., 2020), in a recent study on UK engineering students, articulated that there is more need to implement new reforms to ensure that engineering students should not go down the pathway. Although this figure outs a major challenge, it can be a golden opportunity for stakeholders to rethink and redesign service policies with an effective plan to make engineering education more suitable and accessible for prospective students by reducing stress due to forced transitions. With the majority of urban regions grieved by pandemics, (Gross, 2020) judged small campuses located in rural regions that might be suitable for carrying academic and nonacademic activities where following social distancing protocols can be easier and under control.

*2.4.5 Suitability of an institute under Covid-19*

Suitability under pandemic refers to the institute's efforts and response to making pedagogy available by providing suitable facilities and support services that mitigate the impact of the pandemic on students' education. Literatures representing the suitability of higher educational institutions under the COVID-19 pandemic are almost missing.

## *2.5 Research gap and significance of study*

Many researchers have notable a variety of characteristics influencing institute selection decisions originating from different cultures and economic and social reforms, but all were administered under nonpandemic situations. Many researchers felt that students' characteristics changed during the COVID-19 pandemic, and there is urgency to reposition the framework of policies for the inclusion of diversity, which demanded future research that urges exploring choice characteristics in regard to the suitability and accessibility of engineering institutions during the pandemic.

Moreover, there is no such research to date that provides knowledgeable relationships between students' profiles and their perceptions of institute selection during dangerous COVID-19 pandemics. The importance and timeliness of this study is boundless, as it aimed to explore radical changes that have materialised on choice characteristics and suitability during the calamitous pandemic, COVID-19.

*2.6 Conceptual framework and hypothetical model*

Once students have gathered enough information and undergone human influence, they usually evaluate and prepare prioritized lists of colleges to attend based on their needs and benefits (Kotler & Fox, 1995). This study belongs to the 'choice' stage, which directs the choice decisions of students for a particular engineering institute. For them, the next step is to select their dream engineering institute that is most suited for their personal needs to cope with situational crises arising from the COVID-19 pandemic among the available sets of alternatives.

Human effects that squeeze influences from current students, alumni, friends, family members, school staff and institute staff and students are responsible for making collective decisions for prospective students. The collective decision is based on attractive and beneficial offers made by the institutes, i.e. characteristics in regard to tangible facilities such as location, infrastructure, majors, etc. and intangible services such as teaching-learning, reputation, campus life, etc. These eye-catching characteristics are informed to prospective students through institute promotional activities to create a positive impression in the minds of decision makers. Students and stakeholders capture information from the sources that they think suitable and reliable.

However, during the COVID-19 pandemic, the process of evaluating alternatives involves more intellectual and meticulous screening of institute characteristics in regard to their suitability and accessibility, which has better provisions to deliver educational services by restraining infection and the spread of disease, such as COVID-19. Based on the theoretical and conceptual framework as stated above, the following hypothetical model (Figure 1) will stand for answering research questions. Based on the specified objectives of this study, the conceptual framework and hypothetical model, as demonstrated in Figure 1, are the null hypotheses to be validated based on students' perceptions.

> *Hypothesis one ($H_01$)*
>
> There is no significant difference in the suitability of an engineering institute under the COVID-19 pandemic across their characteristics.
>
> *Hypothesis two ($H_02$)*
>
> There is no significant relationship between the suitability of an engineering institute during the COVID-19 pandemic and the characteristics of that institute.
>
> *Hypothesis three ($H_03$)*

There is no significant relationship between the suitability of an engineering institute during the COVID-19 pandemic and the characteristics of that institute across students' characteristics.

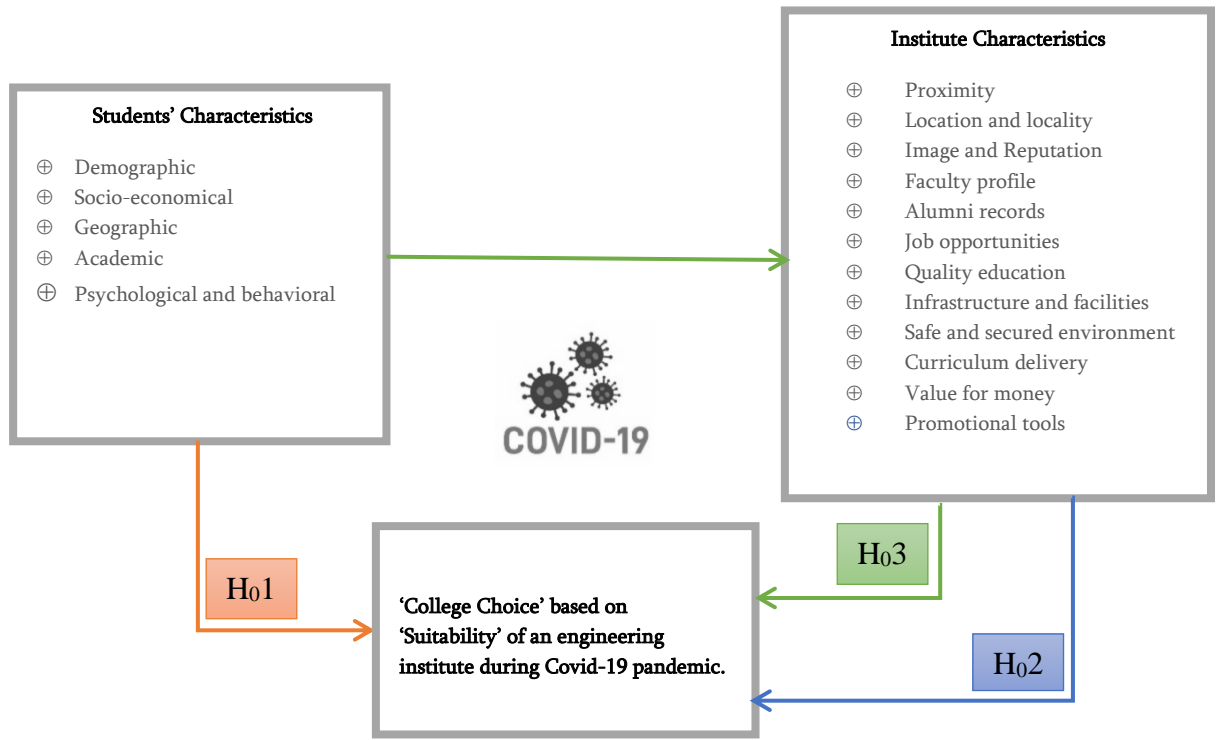

Figure I: Conceptual framework and hypothetical model of engineering college choice

## 3. Research Methodology

### *3.1 Research design*

This study is marketing research about an educational dilemma associated with institute choice of an engineering study, particularly during the COVID-19 pandemic. A literature review aligned with the objective of this study has enabled this study to implement a quantitative method due to its ability to frame hypotheses (Borrego et al., 2009), capabilities to operate on multivariate statistical data (Creswell & Creswell, 2017), ability to analyze relationships with definiteness and transparency (Creswell, 2012b) and reliability (Steckler et al., 1992) and success in educational research (Tight, 2015).

'Student' as primary 'customer', their perceptions must be regularly assessed (Maringe & Gibbs, 1989). The judgment of what students truly receive from service delivery against their expectations is often based on the evaluation of students' perceived experience with a particular service encounter (Yelkur, 2000). Therefore, this study is based on students' perceived experience. Primary data are collected using a survey method that is most suitable as per (Philip Kotler et al., 2016) for collecting preferences and choices from a large number of responses. Experience customers (students who completed their engineering degree) cannot be an appropriate population for this study, as this study is exclusively related to the influence of the

COVID-19 situation, and this population has no experience with their institute choice under the COVID-19 situation. Therefore, students who have just now enrolled in engineering majors were purposefully selected as the study population. Purposive sampling has been chosen decisively because of the knowledge and judgment of researchers (Creswell & Creswell, 2017) (Neuman, 2013), special situations (Neuman, 2013) and investigations of new issues (Etikan et al., 2016) about 'institute choice' during the COVID-19 pandemic.

*3.2 Setting the research scene*

The study is set to report the perceptions of students about their pathway to an engineering degree program during the pandemic. Engineering institutions offering degree courses in engineering and technology situated in the North Maharashtra region of India were chosen as the sampling frame of this study. Admission process under the control of competent authority of Maharashtra State, India for the admission to first year degree engineering program for academic year 2020-2021, the admission process has processed late due to pandemic that started in the month of Dec 2020 and has come to an end on January 31, 2021. Sampling units consisting of student lists representing batches of academic years 2020-201 who recently underwent their institute choice process were chosen from units selected from a sampling frame comprising engineering institutions area wise to cover a variety of demographic, geographical and socioeconomic characteristics of the population. A total of 4300 e-mail addresses of students representing the 2020-2021 academic batch from 69 units of sampling frame (engineering institutions) were collected on request through e-mail during February 2021. To make students more responsive, a self-report survey (Kolb, 2008) was conducted over the internet via the Google Form tool during February 1-15, 2021. During the COVID-19 pandemic, a self-reported survey was very useful, as it avoided the risk of meeting coronavirus by avoiding direct contacts with respondents but at the same time ensured its reach to the expected respondents (students). This method also assisted in receiving responses quickly by providing respondents with better flexibility in time and place and avoided researcher bias.

The survey received 516 responses at a response rate of 12% at the end of Dec 2020. While attempting regression analysis in statistics, (Creswell, 2012a) (Stevens, 2012) recommended a sample size of at least 15 samples per variable, and cases (Floyd & Widaman, 1995) proposed 10 samples per variable. This study included eleven independent variables with 516 valid responses obtained from a self-reported survey. A sample size of 516 for assessing eleven variables, which derives 47 samples per variable, is sufficiently defensible against the traditional arbitrary ratio of 10:1 (Maxwell Scott, 2000).

Purposive sampling and self-administered surveys during the COVID-19 pandemic successfully decoded the purpose of the study and sited the delegate sample that symbolizes the characteristics and profile of the overall population, confirming the knowledge and judgment of the researcher by supplying essential data about choice characteristics during the COVID-19 pandemic situation.

*3.3 Data collection*

A quantitative survey is administered with a list of structured closed-ended questionnaires that were developed on the guidelines provided by (L. Cohen et al., 2007) and (Ary et al., 2010). The questionnaire was initiated with introductory Section I, explaining the purpose and importance of the study and then proceeding further with Sections II and III belonging to students' personal characteristics. Sections IV and V were associated with institute characteristics, and Section VI was related to institute suitability during the COVID-19 pandemic, designed sequentially. Section II consisted of five independent categorical variables that derived answers on the personal characteristics of students; demographic characteristics: gender, socioeconomic characteristics: social class (SES): geographic characteristics: native place, academic characteristics: previous academic performance. Section III presents three independent categorical variables relating to the psychological and behavioral characteristics of students and is designated to acquire students' opinions on priority for curriculum mode and course major. In the next item, the most valuable human influence out of five sub-items in regard to choosing their engineering career was requested. Section IV encompassed eleven continuous variables estimating students' perception to rate on a Likert scale (1 to 5) on characteristics about their dream institute, such as proximity, location and locality, image and reputation, faculty profile, alumni records, job opportunities, quality education, infrastructure and facilities, safety, security, hygienic environment, curriculum delivery and value for money. Last, in Section V, opinions on the most effective promotional tools (independent categorical variable) of their institute were questioned. It comprised eight sub-items: banners and hoarding, institute seminars / presentation, institute website, education fairs, leaflet/brochure, newspapers publicity, radio / TV advertisement and social media from which students were requested to select most impactful and informative promotional tool. In Section VI, appropriateness of their dream institute in delivering engineering education with facilities and services during the COVID-19 pandemic situation was asked to rate on a Likert scale (1 to 5), which is a dependent variable of this study.

Before entering the actual survey, the validity and reliability of the questionnaire was tested through pilot testing (Van Teijlingen & Hundley, 2001) (Kenneth, 2005) with few samples selected from sampling units to understand its language and sequence of questions and was noticed to be suitable for conducting the actual survey.

## 4. Results

To explore the relationships between the suitability of an engineering institute under the COVID-19 situation (CS, dependent and continuous variable), institute characteristics (C1 to C11, independent and continuous variables) and student characteristics (S1 to S8, independent and categorical), SPSS 25 is utilized along with the guidelines suggested by (Julie Pallant, 2020).

*4.1 Statistical tests*

To answer the research question (RQ1) and validate $H_{01}$, ANOVA was conducted as directed by (Powers & Xie, 1999), which demonstrated the association of categorical variables (S1 to S8, student characteristics) with continuous variables (CS, institute characteristics). Regression analysis is the most preferred and useful technique in the case of marketing research pertaining to the study of educational behavior, where the responses are quantitative and continuous (Cohen et al., 2013) (Fox, 2016). Simple regression with continuous variables (C1 to C11, institute characteristics) as predictors was entered simultaneously to identify predictability continuous variables (CS, institute characteristics) to justify RQ2 and $H_{02}$ (Pallant, 2005). After that, to resolve RQ3 and $H_{03}$, the CS variable is analyzed to predict the portion of its variance on account of the variations in C1 to C11 variables across S1 to S8 variables one by one.

*4.2 Statistical fitness of data.*

Before proceeding towards ANOVA and regression analysis, survey responses were tested for various quality measures. To verify the existence of substantial correlation among a set of independent predictors, a multicollinearity test was run using SPSS. The test demonstrated that tolerance (above 0.2) and VIF values (below 10.0) are well within the specified limits and ensured that data have no issue with collinearity and standard errors coupled with the predictors (Daoud, 2018) (Chattefuee & Hadi, 2006) (Miles, 2014). Second, reliability based on internal consistency that validated scales' uniformity with the value of coefficient alpha (Cronbach, 1951) (Churchill Jr, 1979) is reported at a value of 0.941 for the overall model and ranges between 0.916 and 0.937 (>0.6) for all independent predictors. These values express the best fit for this purpose (Nunnally & Bernstein, 1967) (Refer Table II). Based on the corrected item-to-total correlation obtained, as shown in Table II, positive values ranging above 0.3 indicated good internal consistency of the scale (Briggs & Cheek, 1986). Last, Tukey's test was effective in detecting no additivity, which confirmed a sufficient estimate of power.

*4.3 Students' characteristics on suitability under COVID-19 ($H_0 1$)*

Descriptive statistics of eight independent and categorical variables (S1 to S8) representing students' characteristics and results of ANOVA test exhibiting its association with suitability under Covid-19 (CS) is presented in Table I.

To understand students' characteristics precisely and their perception of the suitability of an engineering institute under COVID-19, some characteristics are clubbed to form a group representing similar characteristics. Accordingly, social class (S2) displays higher and lower social class, where higher social class is established by clubbing general and other backward class which showed similar settings in terms of culture and values whereas lower class includes backward class students. Similarly, in regard to the native place of students, the urban residing student group is formed by combining district and taluka residing students, whereas village residing students are termed rural students in accordance with their geographical settings. Likewise, S4, S5, S7 and S8 (refer to Table I) are formed by uniting their similar characters they possessed to form a main group variable that recognizes students' characteristics that are more analogous. This helped in analyzing the relationship clearly, as discussed below.

The results of ANOVA (univariate) showed that gender (S1) did not have any impact on suitability under COVID-19 (CS), as the *F*-value of 2.879 was statistically insignificant at $\rho > 0.05$. This means that there are similar perceptions between male and female groups regarding the suitability of an engineering institute under COVID-19 (CS). Similarly, social class (S2) (*F*-value=0.998, $\rho > 0.05$), native place (S3) (*F*-value=1.052, $\rho > 0.05$), and preinstitute performance (S4) (*F*-value=0.453, $\rho > 0.05$) do not show any statistical significance, indicating that there is similarity across these groups on CS.

**Table I: Suitability under Covid-19 based on Students' characteristics**

| Students Characteristics (S) | Grouped parameters | Suitability under Covid-19 (CS) | | | | |
|---|---|---|---|---|---|---|
| | | N = 516 | µ= 3.64 | *F*-value | ρ | *Hypothesis Support* |
| 1. Gender (S1) | | | | 2.879 | 0.090 | $H_0 1$ |
| Male | | 348 | 3.59 | | | |
| Female | | 168 | *3.76* | | | |
| 2. Social class (S2) | | | | 0.998 | 0.318 | H$_0$1 |
| Higher social class | General class, Other backward class | 393 | *3.96* | | | |
| Lower social class | Backward class | 123 | 3.60 | | | |
| 3. Native place (S3) | | | | 1.052 | 0.306 | H$_0$1 |
| Urban | District, Taluka | 193 | 3.70 | | | |
| Rural | Village | 323 | 3.61 | | | |
| 4. Pre-institute performance (S4) | | | | 0.453 | 0.501 | H$_0$1 |
| High scorer | marks >75% | 210 | 3.68 | | | |
| Low scorer | marks <75% | 306 | 3.62 | | | |
| 5. Major choice (S5) | | | | **8.333** | ***0.004*** | ***H1*** |
| Computer allied courses | Computer and Data Science | 386 | 3.72 | | | |
| Non-computer allied courses | Mechanical, Electrical, Civil and Electronics | 130 | 3.42 | | | |
| 6. Course delivery mode (S6) | | | | **25.601** | ***0.000*** | ***H1*** |
| Online | | 132 | 3.19 | | | |
| Onsite | | 278 | *3.92* | | | |
| Hybrid | | 106 | 3.48 | | | |
| 7. Human influence (S7) | | | | **2.686** | ***0.046*** | ***H1*** |
| Institute | Institute staff and institute student | 225 | 3.69 | | | |
| Community | Friends and relatives | 150 | 3.52 | | | |
| Family | Family members | 90 | *3.84* | | | |
| Pre-institute school | Pre-institute school | 51 | 3.43 | | | |
| 8. Information source (S8) | | | | **2.560** | ***0.049*** | ***H1*** |
| Non-interactive and offline | Media advt., banners, hoardings, leaflets | 81 | 3.67 | | | |
| Non-interactive and online | Website | 162 | *3.77* | | | |
| Interactive and offline | Seminars, exhibitions, school visits | 114 | 3.71 | | | |
| Interactive and online | Social media | 159 | 3.46 | | | |

Source: ANOVA run through SPSS
Note: Values in bold and italic are significant

When institute suitability under COVID-19 (CS) is assessed with students' major choice (S5), significant differences are observed across the groups ($F$-value=8.333, $\rho$=0.004 <0.005): students who selected computer allied majors and noncomputer majors. The students who opted computer-related major courses observed having greater attention to CS ($\mu$=3.72) than the students who chose noncomputer majors. With an $F$-value of 25.601 at $\rho$=<0.0001, the course delivery mode (S6) has the greatest effect on suitability under the COVID-19 situation. CS seems to be the greatest concern for those preferring onsite learning ($\mu$=3.92) and hybrid learning ($\mu$=3.48) than the students' highlighting online learning ($\mu$=3.19). There was a significant difference between CS and prime human influence (S7) responsible for students' institute choice, with the association counting an $F$-value of 2.686 at $\rho$=<0.05. Family has greater influence ($\mu$=3.84), followed by institute representatives ($\mu$=3.69), like staff and students who have provided required information about suitability of an engineering institute under Covid-19 than other human influences determining their inclusion into engineering institute. Information sources (S8) are spotted to have statistical significance with CS that affected students' institute choice with an $F$-value of 2.560 at $\rho$=<0.05. A non-interactive and online information source, website ($\mu$=3.77), is proven to have the greatest impact in providing information about CS that favoured their institute choice, followed by interactive and offline

($\mu$=3.71) means such as face-to-face communications encompassing counselling seminars and school contacts. Interactive and offline sources and social media ($\mu$=3.46), however, had a lesser impact than other information sources in providing knowledge about suitability under COVID-19, which supports institute choice decisions.

Hence, the null hypothesis ($H_0 1$) is rejected for S5, S6, S7 and S8, where there is a significant difference noted with CS. Alternative hypothesis H1 is accepted in this case. In contrast, the null hypothesis ($H_0 1$) is retained for S1, S2, S3 and S4, where a significant difference does not exist with CS.

*4.4 Institute characteristics on suitability under COVID-19 ($H_0 2$)*

Table II presents descriptive statistics of students' perceptions about institute characteristics that are responsible for governing institute choice, along with squared multiple correlation ($R^2$) and Cronbach's α. The overall Cronbach's α for all institute characteristics (C1 to C11 and CS) is 0.941, indicating strong strength about scale uniformity and reliability towards measuring institute choice. Image and reputation (C3) ($\mu$=4.140, $R^2$=0.718), followed by location and locality (C2) ($\mu$=4.120, $R^2$=0.621), quality education (C7) ($\mu$=4.080, $R^2$=0.756), safe and secured environment (C9) ($\mu$=4.080, $R^2$=0.686) and campus placement (C6) ($\mu$=4.070, $R^2$=0.663), are among the top five institute characteristics that have enchanted institute choice decisions for the students. However, proximity ($\mu$=3.480, $R^2$=0.297) and suitability under COVID-19 ($\mu$=3.640, $R^2$=0.435) are at the bottom of the list and constitute important characteristics controlling institute choice decisions for students.

**Table II: Descriptive statistics and reliability of institute characteristics**

| Institute characteristics | Code | Mean ($\mu$) | Standard Deviation ($\sigma$) | N | Corrected Item-Total $R$ | $R^2$ | Cronbach's α if Item Deleted |
|---|---|---|---|---|---|---|---|
| Group Cronbach's α= 0.941 | | | | | | | |
| Proximity | C1 | 3.480 | 1.321 | 516 | 0.346 | 0.297 | 0.937 |
| Location and locality | C2 | 4.120 | 0.826 | 516 | 0.719 | 0.621 | 0.920 |
| Image and reputation | C3 | 4.140 | 0.784 | 516 | 0.778 | 0.718 | 0.918 |
| Faculty profile | C4 | 4.060 | 0.794 | 516 | 0.805 | 0.715 | 0.917 |
| Alumni profile | C5 | 4.020 | 0.851 | 516 | 0.766 | 0.654 | 0.918 |
| Campus placements | C6 | 4.070 | 0.812 | 516 | 0.753 | 0.663 | 0.919 |
| Quality education | C7 | 4.080 | 0.774 | 516 | 0.840 | 0.756 | 0.916 |
| Infrastructure faculty | C8 | 4.020 | 0.785 | 516 | 0.765 | 0.628 | 0.919 |
| Safe and secured environment | C9 | 4.080 | 0.760 | 516 | 0.798 | 0.686 | 0.918 |
| Curriculum delivery | C10 | 4.020 | 0.797 | 516 | 0.836 | 0.761 | 0.916 |
| Value for money | C11 | 3.870 | 0.928 | 516 | 0.707 | 0.539 | 0.920 |
| Covid-19 suitability | CS | 3.640 | 1.049 | 516 | 0.632 | 0.435 | 0.924 |

Source: Reliability test run through SPSS

**Table III: Regression analysis CS←(C1 to C11)**

Dependable variable - Suitability under Covid-19 (CS)

| Independent variables (predictors) | Unstandardized Coefficients | | Standardized Coefficients | $t$ | $\rho$ | Collinearity Statistics | | Hypothesis Support |
|---|---|---|---|---|---|---|---|---|
| | $B$ | Std. Error | $\beta$ | | | Tolerance | VIF | |
| (Constant) | -0.363 | 0.221 | | -1.644 | 0.101 | | | |
| C1 | 0.177 | 0.028 | 0.223 | **6.265** | **0.000** | 0.885 | 1.129 | **H2** |
| C2 | -0.035 | 0.069 | -0.027 | -0.503 | 0.615 | 0.379 | 2.637 | $H_02$ |
| C3 | 0.191 | 0.084 | 0.143 | **2.281** | **0.023** | 0.285 | 3.510 | **H2** |
| C4 | 0.008 | 0.083 | 0.006 | 0.094 | 0.925 | 0.285 | 3.505 | $H_02$ |
| C5 | -0.025 | 0.070 | -0.021 | -0.361 | 0.719 | 0.346 | 2.889 | $H_02$ |
| C6 | -0.030 | 0.075 | -0.023 | -0.401 | 0.688 | 0.337 | 2.970 | $H_02$ |
| C7 | 0.363 | 0.090 | 0.268 | **4.017** | **0.000** | 0.252 | 3.966 | **H2** |
| C8 | 0.037 | 0.073 | 0.028 | 0.502 | 0.616 | 0.372 | 2.690 | $H_02$ |
| C9 | 0.050 | 0.082 | 0.036 | 0.601 | 0.548 | 0.314 | 3.185 | $H_02$ |
| C10 | 0.186 | 0.090 | 0.141 | **2.073** | **0.039** | 0.241 | 4.144 | **H2** |
| C11 | 0.091 | 0.056 | 0.081 | 1.638 | 0.102 | 0.463 | 2.158 | $H_02$ |

| Regression Model | | | | ANOVA[a] | | Hypothesis support |
|---|---|---|---|---|---|---|
| $R$ | $R^2$ | Adjusted $R^2$ | SE | F-value | $p$ | |
| 0.659[a] | 0.435 | 0.422 | 0.797 | 35.220 | 0.000[b] | **H2** |

Source: Regression analysis run through SPSS
Note: Values in bold and italic are significant. a: dependent variable (CS), b: predictors (C1 to C11)

The regression model (refer to Table III) shows that the relationship between eleven independent characteristics of institutions (C1 to C11) and the dependent variable (CS) is statistically significant ($F$-value=35.220, p<0.000) in predicting the relationship CS←(C1 to C11). It further confirmed that the model is associated with a large coefficient of correlation ($R$=0.659) (Cohen, 1988) and coefficient of regression ($R^2$=0.435) having moderate to strong strength (Chin, 1998) (Glenn & Shelton, 1983) (Cohen et al., 2013) in predicting CS, even though the data narrate unpredictable human behavior. This model further guaranteed that CS significantly predicted 43.5% of its variance accounted for by independent predictors (C1 to C11). It is also observed that institutional characteristics, i.e. suitability under Covid-19 (CS) is statistically significant and is in positive relationship with proximity (C1) ($B$=0.177, $\beta$=0.223, t=6.265, $\rho$<0.000), institute image and reputation (C3) ($B$=0.191, $\beta$=0.143, t=2.281, $\rho$<0.05), quality education (C7) ($B$=0.363, $\beta$=0.268, t=4.017, $\rho$<0.000)) and curriculum delivery (C10) ($B$=0.186, $\beta$=0.141, t=2.073, $\rho$<0.05) that are offered by their institute. Unstandardized estimates ($B$) revealed that when C1, C3, C7 and C10 increase by one unit, CS increases by 0.177, 0.191, 0.363, and 0.186 units, respectively, while keeping other variables stable. On the other hand, C2, C4, C5, C6, C8, C9 and C11 had no statistical relationship with CS.

There was statistical significance in predicting CS on account for by independent variables ($F$=35.220, $p$<0.000). The null hypothesis ($H_0 2$) is rejected in cases where the independent variables C1, C3, C7 and C10 are statistically related to the dependent variable (CS). Therefore, alternative hypothesis H2 is accepted. On the other hand, the null hypothesis ($H_0 2$) is retained for C2, C4, C5, C6, C8, C9 and C11, where a statistical association does not exist with CS.

*4.5 Institute characteristics on suitability under COVID-19 across students' characteristics ($H_0 3$)*

To test the relationship between institute characteristics (C1 to C11) and suitability under the COVID-19 pandemic (CS), i.e. CS←(C1 to C11) across students' characteristics (S1 to S8). Data were split based on categorical variables (student characteristics) with the split file option available in SPSS, and then simple regression was performed as executed in the previous case. The statistical results are discussed below (Refer Table IV).

Referring to gender, the model CS←(C1 to C11) for males ($R=0.66$, $R^2=0.44$, $F=23.961$, $p<0.000$) and females ($R=0.70$, $R^2=0.49$, $F=13.394$, $p<0.000$) fits a similar way. CS is significantly predicted by C1, C3, C7 and C10 for males, while C1 and C4 are significant estimates for females. In the case of the social class of students, the model was better suited to higher social classes ($R=0.81$, $R^2=0.66$, $F=19.632$, $p<0.000$) than lower social classes ($R=0.62$, $R^2=0.39$, $F=21.899$, $p<0.000$). C1, C3 and C7 for the higher social class and C1, C4 and C7 for the lower class contributed significantly to CS. Across the native place of students, CS contributed equally by the effect of C1, C6, C7 and C9 for urban students ($R=0.71$, $R^2=0.50$, $F=16.553$, $p<0.000$) and C1, C7 and C11 for rural students ($R=0.66$, $R^2=0.43$, $F=21.569$, $p<0.000$). CS is accounted for by nearly the same value of estimates of $R$ and $R^2$. Across the preinstitute performance group, C1, C2, C7 and C10 were significant contributors to high scores ($R=0.67$, $R^2=0.45$, $F=14.913$, $p<0.000$), and C1, C7 and C11 were significant contributors to low scores ($R=0.68$, $R^2=0.46$, $F=22.443$, $p<0.000$).

In the case of major choice, the regression model CS←(C1 to C11) is more appropriate for students adopting non-computer allied majors ($R=0.77$, $R^2=0.60$, $F=15.921$, $p<0.000$) than for students accepting computer allied majors ($R=0.65$, $R^2=0.42$, $F=24.982$, $p<0.000$). C1, C2, C3, C7, C9 and C19 and C1, C2, C7, C9 and C10 are the key significant influencers in predicting CS for students adopting computer allied majors and noncomputer allied majors, respectively. Understanding delivery mode, the regression model is more bearing on online ($R=0.70$, $R^2=0.49$, $F=10.642$, $p<0.000$) and hybrid ($R=0.70$, $R^2=0.49$, $F=8.267$, $p<0.000$) than onsite mode ($R=0.66$, $R^2=0.44$, $F=19.011$, $p<0.000$). Only two significant predictors, C1 and C11, govern CS in the case of the online mode. CS is estimated by virtue of C1, C7 C10 and C11 and C1, C3 and C7 for hybrid and onsite delivery modes, respectively.

The regression model CS←(C1 to C11) for information source have shown that it suits better for the non-interacting and online source (website) ($R=0.77$, $R^2=0.59$, $F=13.126$, $p<0.000$), and interactive and offline (face-to-face counselling) ($R=0.70$, $R^2=0.49$, $F=13.422$, $p<0.000$), then non-interactive print and media advertisement ($R=0.66$, $R^2=0.44$, $F=4.868$, $p<0.000$) and interactive and online sources (social media) ($R=0.63$, $R^2=0.40$, $F=8.744$, $p<0.000$). C1, C2, C4 and C11 contribute significantly to websites, C1 and C7 estimate significantly for print/advertisement media, C1 and C10 significantly affect face-to-face interaction and C1, C7, C11 significantly account for social media on CS.

In view of human influence, the model better poised for school ($R=0.74$, $R^2=0.55$, $F=4.390$, $p<0.000$), for family ($R=0.72$, $R^2=0.51$, $F=7.451$, $p<0.000$), for institute representative ($R=0.71$, $R^2=0.50$, $F=19.333$, $p<0.000$) than community ($R=0.65$, $R^2=0.42$, $F=9.052$, p$<0.000$). CS was significantly accounted for by C1, C3, C7, C8 and C11 for institute representatives, C1 and C10 for communities, and C1 and C7 for families. For school, C6, C7 and C11 are the main predictors in assessing CS; however, they are not statistically significant.

There was a statistically significant relationship found in predicting the dependent variable (CS) on account for by independent variables across every student's characteristic. Hence, the null hypothesis (H$_0$3) is rejected, and alternative hypothesis H3 is accepted (Refer Table IV).

**Table IV: Regression analysis CS←(C1 to C11) across students' characteristics**

| | C1 | C2 | C3 | C4 | C5 | C6 | C7 | C8 | C9 | C10 | C11 | $R$ | $R^2$ | Adjusted $R^2$ | SE | $F$-value | $p$ | Hypothesis support |
|---|---|---|---|---|---|---|---|---|---|---|---|---|---|---|---|---|---|---|
| **Gender** | | | | | | | | | | | | | | | | | | |
| Male | | | | | | | | | | | | 0.66 | 0.44 | 0.42 | 0.85 | 23.961 | 0.000 | H3 |
| β | 0.22 | 0.01 | 0.20 | -0.09 | -0.06 | -0.01 | 0.28 | 0.02 | -0.03 | 0.19 | 0.11 | | | | | | | |
| t | 5.09 | 0.16 | 2.60 | -1.19 | -0.92 | -0.09 | 3.49 | 0.35 | -0.35 | 2.26 | 1.76 | | | | | | | |
| p | **0.00** | 0.88 | **0.01** | 0.24 | 0.36 | 0.93 | **0.00** | 0.73 | 0.73 | **0.02** | 0.08 | | | | | | | |
| Female | | | | | | | | | | | | 0.70 | 0.49 | 0.45 | 0.66 | 13.394 | 0.000 | H3 |
| β | 0.22 | -0.10 | 0.00 | 0.28 | 0.01 | -0.11 | 0.16 | 0.13 | 0.11 | 0.15 | 0.03 | | | | | | | |
| t | 3.39 | -0.86 | 0.01 | 2.59 | 0.12 | -1.08 | 1.29 | 1.24 | 1.09 | 1.24 | 0.39 | | | | | | | |
| p | **0.00** | 0.39 | 0.99 | **0.01** | 0.91 | 0.28 | 0.20 | 0.22 | 0.28 | 0.22 | 0.70 | | | | | | | |
| **Social Class** | | | | | | | | | | | | | | | | | | |
| Higher social class | | | | | | | | | | | | 0.81 | 0.66 | 0.63 | 0.62 | 19.632 | 0.000 | H3 |
| β | 0.25 | -0.02 | 0.21 | -0.11 | -0.03 | -0.01 | 0.21 | 0.09 | 0.02 | 0.14 | 0.07 | | | | | | | |
| t | 5.91 | -0.32 | 2.78 | -1.44 | -0.49 | -0.18 | 2.86 | 1.28 | 0.31 | 1.92 | 1.28 | | | | | | | |
| p | **0.00** | 0.75 | **0.01** | 0.15 | 0.62 | 0.86 | **0.00** | 0.20 | 0.76 | 0.06 | 0.20 | | | | | | | |
| Lower social class | | | | | | | | | | | | 0.62 | 0.39 | 0.37 | 0.84 | 21.899 | 0.000 | H3 |
| β | 0.16 | 0.03 | -0.03 | 0.26 | -0.03 | -0.05 | 0.54 | -0.13 | -0.01 | 0.18 | 0.03 | | | | | | | |
| t | 2.61 | 0.39 | -0.25 | 2.22 | -0.26 | -0.51 | 3.32 | -1.30 | -0.09 | 1.10 | 0.25 | | | | | | | |
| p | **0.01** | 0.70 | 0.80 | **0.03** | 0.80 | 0.61 | **0.00** | 0.20 | 0.93 | 0.28 | 0.81 | | | | | | | |
| **Native Place** | | | | | | | | | | | | | | | | | | |
| Urban | | | | | | | | | | | | | | | | | | |
| β | 0.16 | 0.02 | 0.01 | 0.13 | -0.06 | -0.27 | 0.32 | 0.20 | 0.21 | 0.12 | 0.02 | 0.71 | 0.50 | 0.47 | 0.74 | 16.553 | 0.000 | H3 |
| t | 2.64 | 0.20 | 0.07 | 1.35 | -0.71 | -2.61 | 3.47 | 1.93 | 2.14 | 1.21 | 0.34 | | | | | | | |
| p | **0.01** | 0.84 | 0.95 | 0.18 | 0.48 | **0.01** | **0.00** | 0.06 | **0.03** | 0.23 | 0.74 | | | | | | | |
| Rural | | | | | | | | | | | | 0.66 | 0.43 | 0.41 | 0.82 | 21.569 | 0.000 | H3 |
| β | 0.25 | -0.04 | 0.15 | 0.01 | 0.02 | 0.03 | 0.34 | -0.08 | -0.08 | 0.11 | 0.14 | | | | | | | |

|  |  |  |  |  |  |  |  |  |  |  |  |  |  |  |  |  |  |
|---|---|---|---|---|---|---|---|---|---|---|---|---|---|---|---|---|---|
| *t* | 5.58 | -0.67 | 1.83 | 0.09 | 0.24 | 0.40 | 3.30 | -1.09 | -0.96 | 1.20 | 2.03 |  |  |  |  |  |  |
| *p* | **0.00** | 0.50 | 0.07 | 0.93 | 0.81 | 0.69 | **0.00** | 0.28 | 0.34 | 0.23 | **0.04** |  |  |  |  |  |  |

**Pre-institute performance**

| | | | | | | | | | | | | | | | | | |
|---|---|---|---|---|---|---|---|---|---|---|---|---|---|---|---|---|---|
| High Scorer | | | | | | | | | | | | 0.67 | 0.45 | 0.42 | 0.77 | 14.913 | 0.000 | H3 |
| β | 0.34 | -0.18 | 0.17 | -0.10 | 0.01 | 0.07 | 0.30 | -0.09 | 0.00 | 0.35 | -0.03 | | | | | | | |
| *t* | 6.04 | -2.21 | 1.69 | -0.92 | 0.06 | 0.72 | 2.80 | -0.87 | 0.03 | 3.29 | -0.40 | | | | | | | |
| *p* | **0.00** | **0.03** | 0.09 | 0.36 | 0.95 | 0.47 | **0.01** | 0.39 | 0.98 | **0.00** | 0.69 | | | | | | | |
| Low Scorer | | | | | | | | | | | | 0.68 | 0.46 | 0.44 | 0.81 | 22.443 | 0.000 | H3 |
| β | 0.16 | 0.07 | 0.12 | 0.06 | 0.00 | -0.10 | 0.25 | 0.06 | 0.07 | 0.04 | 0.14 | | | | | | | |
| *t* | 3.51 | 0.95 | 1.42 | 0.71 | 0.05 | -1.31 | 2.90 | 0.89 | 0.79 | 0.47 | 2.18 | | | | | | | |
| *p* | **0.00** | 0.34 | 0.16 | 0.48 | 0.96 | 0.19 | **0.00** | 0.38 | 0.43 | 0.64 | **0.03** | | | | | | | |

**Major Choice**

| | | | | | | | | | | | | | | | | | |
|---|---|---|---|---|---|---|---|---|---|---|---|---|---|---|---|---|---|
| Computer allied | | | | | | | | | | | | 0.65 | 0.42 | 0.41 | 0.79 | 24.982 | 0.000 | H3 |
| β | 0.25 | -0.16 | 0.17 | -0.10 | 0.01 | -0.15 | 0.25 | 0.12 | 0.17 | 0.07 | 0.21 | | | | | | | |
| *t* | 6.00 | -2.43 | 2.37 | -1.36 | 0.17 | -1.83 | 3.23 | 1.80 | 2.28 | 0.81 | 3.20 | | | | | | | |
| *p* | **0.00** | **0.02** | **0.02** | 0.17 | 0.86 | 0.07 | **0.00** | 0.07 | **0.02** | 0.42 | **0.00** | | | | | | | |
| Non-Computer allied | | | | | | | | | | | | 0.77 | 0.60 | 0.56 | 0.73 | 15.921 | 0.000 | H3 |
| β | 0.25 | 0.26 | 0.04 | 0.15 | 0.02 | 0.03 | 0.38 | -0.16 | -0.22 | 0.25 | -0.03 | | | | | | | |
| *t* | 3.59 | 2.92 | 0.35 | 1.21 | 0.18 | 0.34 | 2.65 | -1.84 | -2.30 | 2.09 | -0.37 | | | | | | | |
| *p* | **0.00** | **0.00** | 0.73 | 0.23 | 0.85 | 0.73 | **0.01** | 0.07 | **0.02** | **0.04** | 0.71 | | | | | | | |

**Delivery mode**

| | | | | | | | | | | | | | | | | | |
|---|---|---|---|---|---|---|---|---|---|---|---|---|---|---|---|---|---|
| Online | | | | | | | | | | | | 0.70 | 0.49 | 0.45 | 0.87 | 10.642 | 0.000 | H3 |
| β | 0.35 | -0.01 | 0.07 | -0.05 | 0.04 | 0.02 | -0.09 | 0.20 | -0.04 | 0.15 | 0.29 | | | | | | | |
| *t* | 4.91 | -0.09 | 0.44 | -0.36 | 0.26 | 0.17 | -0.55 | 1.55 | -0.26 | 0.93 | 2.75 | | | | | | | |
| *p* | **0.00** | 0.93 | 0.66 | 0.72 | 0.80 | 0.87 | 0.58 | 0.12 | 0.80 | 0.36 | **0.01** | | | | | | | |
| Hybrid | | | | | | | | | | | | 0.70 | 0.49 | 0.43 | 0.70 | 8.267 | 0.000 | H3 |
| | β | 0.21 | -0.21 | 0.05 | 0.22 | -0.13 | 0.16 | 0.53 | -0.21 | 0.01 | 0.40 | -0.27 | | | | | | |
| | t | 2.58 | -1.47 | 0.36 | 1.55 | -1.01 | 1.19 | 3.73 | -1.57 | 0.08 | 3.03 | -2.16 | | | | | | |
| | p | **0.01** | 0.15 | 0.72 | 0.12 | 0.31 | 0.24 | **0.00** | 0.12 | 0.94 | **0.00** | **0.03** | | | | | | |
| Onsite | | | | | | | | | | | | | 0.66 | 0.44 | 0.42 | 0.72 | 19.011 | 0.000 | H3 |
| | β | 0.16 | 0.01 | 0.22 | -0.02 | 0.05 | -0.10 | 0.33 | 0.07 | 0.00 | 0.11 | 0.04 | | | | | | |
| | t | 3.27 | 0.07 | 2.61 | -0.27 | 0.68 | -1.15 | 3.40 | 0.98 | -0.04 | 1.12 | 0.58 | | | | | | |
| | p | **0.00** | 0.94 | **0.01** | 0.79 | 0.50 | 0.25 | **0.00** | 0.33 | 0.97 | 0.27 | 0.56 | | | | | | |
| **Information Source** | | | | | | | | | | | | | | | | | | | |
| Non-interactive and offline | | | | | | | | | | | | | 0.66 | 0.44 | 0.35 | 0.84 | 4.868 | 0.000 | H3 |
| | β | 0.26 | -0.07 | 0.20 | 0.10 | 0.12 | -0.28 | 0.48 | 0.22 | 0.01 | -0.12 | -0.12 | | | | | | |
| | t | 2.65 | -0.39 | 1.02 | 0.50 | 0.57 | -1.36 | 2.90 | 1.26 | 0.05 | -0.61 | -1.04 | | | | | | |
| | p | **0.01** | 0.70 | 0.31 | 0.62 | 0.57 | 0.18 | **0.01** | 0.21 | 0.96 | 0.54 | 0.30 | | | | | | |
| Non-interactive and online | | | | | | | | | | | | | 0.70 | 0.49 | 0.45 | 0.73 | 13.126 | 0.000 | H3 |
| | β | 0.18 | 0.31 | 0.15 | -0.27 | -0.01 | 0.04 | 0.26 | -0.08 | 0.09 | -0.02 | 0.21 | | | | | | |
| | t | 2.86 | 2.98 | 1.45 | -2.48 | -0.09 | 0.33 | 1.80 | -0.88 | 0.84 | -0.18 | 2.62 | | | | | | |
| | p | **0.01** | **0.00** | 0.15 | **0.01** | 0.93 | 0.75 | 0.07 | 0.38 | 0.40 | 0.86 | **0.01** | | | | | | |
| Interactive and offline | | | | | | | | | | | | | 0.77 | 0.59 | 0.55 | 0.82 | 13.422 | 0.000 | H3 |
| | β | 0.15 | -0.08 | 0.18 | 0.19 | 0.00 | -0.06 | 0.16 | 0.01 | -0.03 | 0.49 | -0.17 | | | | | | |
| | t | 2.04 | -0.83 | 1.11 | 1.07 | 0.01 | -0.50 | 0.95 | 0.09 | -0.28 | 2.79 | -1.24 | | | | | | |
| | p | **0.04** | 0.41 | 0.27 | 0.29 | 0.99 | 0.62 | 0.34 | 0.93 | 0.78 | **0.01** | 0.22 | | | | | | |
| Interactive and online | | | | | | | | | | | | | 0.63 | 0.40 | 0.35 | 0.78 | 8.744 | 0.000 | H3 |
| | β | 0.27 | -0.12 | -0.01 | 0.10 | -0.09 | -0.01 | 0.39 | -0.06 | -0.04 | 0.15 | 0.22 | | | | | | |
| | t | 3.93 | -1.01 | -0.10 | 0.97 | -0.80 | -0.06 | 3.04 | -0.54 | -0.32 | 1.07 | 2.11 | | | | | | |
| *p* | **0.00** | 0.32 | 0.92 | 0.33 | 0.43 | 0.95 | **0.00** | 0.59 | 0.75 | 0.29 | **0.04** | | | | | | | |
| **Human Influence** | | | | | | | | | | | | | | | | | | |
| Institute | | | | | | | | | | | | 0.71 | 0.50 | 0.47 | 0.78 | 19.333 | 0.000 | H3 |
| β | 0.18 | 0.00 | 0.19 | -0.07 | -0.03 | -0.05 | 0.27 | 0.20 | -0.10 | 0.10 | 0.19 | | | | | | | |
| *t* | 3.22 | 0.06 | 2.20 | -0.79 | -0.33 | -0.70 | 3.02 | 2.53 | -1.14 | 0.99 | 2.59 | | | | | | | |
| *p* | **0.00** | 0.95 | **0.03** | 0.43 | 0.75 | 0.48 | **0.00** | **0.01** | 0.26 | 0.32 | **0.01** | | | | | | | |
| Community | | | | | | | | | | | | 0.65 | 0.42 | 0.37 | 0.85 | 9.052 | 0.000 | H3 |
| β | 0.27 | -0.16 | 0.11 | 0.05 | -0.11 | 0.07 | 0.20 | -0.19 | 0.17 | 0.44 | -0.05 | | | | | | | |
| *t* | 3.99 | -1.45 | 0.81 | 0.37 | -1.10 | 0.55 | 1.51 | -1.68 | 1.35 | 3.10 | -0.59 | | | | | | | |
| *p* | **0.00** | 0.15 | 0.42 | 0.72 | 0.27 | 0.58 | 0.13 | 0.10 | 0.18 | **0.00** | 0.55 | | | | | | | |
| Family | | | | | | | | | | | | 0.72 | 0.51 | 0.44 | 0.67 | 7.451 | 0.000 | H3 |
| β | 0.36 | 0.11 | -0.16 | 0.13 | 0.31 | -0.36 | 0.55 | -0.09 | 0.20 | -0.17 | -0.04 | | | | | | | |
| *t* | 3.84 | 0.68 | -0.83 | 0.70 | 1.45 | -1.77 | 2.01 | -0.57 | 1.04 | -1.07 | -0.25 | | | | | | | |
| *p* | **0.00** | 0.50 | 0.41 | 0.48 | 0.15 | 0.08 | **0.05** | 0.57 | 0.30 | 0.29 | 0.80 | | | | | | | |
| School | | | | | | | | | | | | 0.74 | 0.55 | 0.43 | 0.82 | 4.390 | 0.000 | H3 |
| β | 0.19 | -0.44 | 0.27 | -0.24 | 0.05 | 0.47 | 0.35 | 0.07 | 0.20 | -0.35 | 0.34 | | | | | | | |
| *t* | 1.66 | -1.13 | 0.68 | -0.77 | 0.19 | 1.47 | 0.82 | 0.17 | 0.91 | -1.09 | 1.87 | | | | | | | |
| *p* | 0.11 | 0.27 | 0.50 | 0.45 | 0.85 | 0.15 | 0.42 | 0.87 | 0.37 | 0.28 | 0.07 | | | | | | | |

Source: Regression analysis run through SPSS
Note: Values in bold are significant. dependent variable is CS, predictors are C1 to C11

## 5. Statistical inference and discussions

### 5.1 Institute choice characteristics

This study has verified choice characteristics in regard to the selection of an EI. Out of twelve institute characteristics, eleven traditional institutes' characteristics and one new characteristic, suitability under COVID-19, endorsed due to the pandemic situation, have shown important contributions in deciding EI choices. The choice process for an engineering EI has thus safeguarded students' mobility within the domestic market, at least for Indian EIs. Despite the fact that choice characteristics in the pandemic acted in a comparative manner to ordinary conditions, they incredibly affected the suitability of an EI under the COVID-19 pandemic situation.

Image and reputation ($\mu$=4.140, $R^2$=0.718) are rated of highest importance for the overall sample in making their EI choice, similar to most of the literature, such as (S. Briggs, 2006) (M. Palmer et al., 2004) (Conard & Conard, 2000) (Gill et al., 2018). Next, location and locality ($\mu$=4.120, $R^2$=0.621), as the measures of EI choice (Weisser, 2020) as evidenced by this study, has been referred to as accessibility and suitability of hi-tech facilities and amenities wherein it is situated (Hannagan, 1992; Kotler & Fox, 1995). Curriculum delivery ($\mu$=4.020, $R^2$=0.761), campus placement activities ($\mu$=4.070, $R^2$=0.663), faculty profile ($\mu$=4.060, $R^2$=0.715), alumni profile ($\mu$=4.020, $R^2$=0.654), and quality education ($\mu$=4.080, $R^2$=0.663) are noted to be the foremost selection criteria in EI choice. Providing a safe and secured campus ($\mu$=4.080, $R^2$=0.686) has also contributed to the list of important EI characteristics, and choice now a days has articulated to be most critical in fetching students on campus (Calitz et al., 2020). Infrastructure and facilities ($\mu$=4.020, $R^2$=0.628) provided by the EI govern students' choice because EI (I. Price et al., 2003) is also applicable here. These top dominating characteristics have been shown to be important in making EI choices in various studies (Hemsley-Brown & Oplatka, 2015c) pertaining to different cultures regularly. (Davies & Guppy, 1997) revealed similar results regarding the image and reputation of engineering institutions.

Surprisingly, during pandemic circumstances, this study stayed diverted about proximity, value for money and suitability under COVID-19, which are notorious measures in such situations. Proximity ($\mu$=3.480, $R^2$=0.297), value for money ($\mu$=3.870, $R^2$=0.539) and suitability under COVID-19 ($\mu$=3.640, $R^2$=0.435) are at the bottom of the list of choice measures. In a COVID-19 situation, while deciding an EI of their choice, students have overlooked nearness to their hometown, effectiveness in regard to efforts, cost and time, and aptness about virus protection. This means that students' attitudes towards EI are positive, wherein they have enrolled due to the top five stunning characteristics, as discussed above.

### 5.2 Students' characteristics on suitability under pandemic

The ANOVA test demonstrated statistical insignificance on students' perceptions of the suitability of an EI under COVID-19 across their demographic: gender, socioeconomic: social class, geographical: native place, and academic: preinstitute school performance. This suggests that the importance of suitability under COVID-19 bears almost equal weightage across these groups, as the pandemic is hampering the higher education of all students all way (Marinoni et al., 2020); students may be male or female, belong to higher social or lower social classes, reside in urban or rural areas, and have high or low scores.

Major choice did have an impact on suitability under COVID-19 ($F$-value=8.333, $p<0.005$). Students who selected allied computer courses ($\mu=3.72$) as their major perceived greater importance to suitability under the COVID-19 pandemic than students with non-computer allied majors ($\mu=3.42$). Until now, in India, delivery of higher education during pandemic situations has been performed online with students sitting and performing at home with their computer/laptop/mobile and discovered effective in HE delivery (Mae, 2020). Subjects covered in computer allied majors can be learned and understood at home with a hi-tech online platform. This made students enter computer allied majors to perceive greater suitability under COVID-19. On the other hand, non-computer allied majors are difficult to learn and understand from home (CDC, 2021), unfavourable for online learning (Zia, 2020a) (K. A. Bird, 2020) and can further bring academic loss (Najmul Hasan & Bao, 2020) during the pandemic. Therefore, these students perceive lower suitability under COVID-19 than students associated with computer allied majors.

Course delivery has shown a significant impact on suitability under COVID-19 ($F$-value=25.601, $p<0.000$), having a greater effect for onsite ($\mu=3.92$) and hybrid ($\mu=3.48$) than online delivery ($\mu=3.19$). Recently, in the Indian context, (Kundu & Bej, 2021) discovered that online learning during the pandemic is not suitable due to fear, insecurities, and a number of challenges associated with digital connections. Online delivery of engineering may create a skills gap (Chadha et al., 2020) and may lead to academic loss (Najmul Hasan & Bao, 2020). The hybrid delivery mode, a mixed version associated with online theory pedagogy and onsite pedagogy for practical and lab work, with safe distancing measures has proven to be successful (Tan, 2020) after the reopening of EIs during the pandemic. Onsite pedagogy is better for learning and understanding tough and hard subjects in person, and learning them from home (online) might be difficult (CDC, 2021) and unfavorable (Zia, 2020) (Bird, 2020), which can bring academic loss (Hasan & Bao, 2020) during pandemics. Because of this, engineering students yearned for onsite and hybrid pedagogy during the COVID-19 pandemic by following social distancing measures, which is appropriate by (Gurukkal, 2020).

The suitability of EI under COVID-19 ($F$-value=2.686, $p<0.05$) is significantly associated with human influence, which is supportive of EI choice decisions. Family as a close credible source of advice (Wong et al., 2019) and institute staff and students as a direct source (Veloutsou et al., 2005a) are substantial reliable sources in guiding and providing superior information about the suitability of EI under pandemic conditions. Furthermore, their assistance in this regard has directed students' pathway into an institute of their choice, which is found to be in accord with the findings of (María Cubillo-Pinilla et al., 2006). Schools, friends, and relatives may perhaps not be in position to consult students about the suitability of EI under pandemic accurately due to non-availability of information about institute actions during pandemic situation. Therefore, students who perceived them as their key influencers in EI choice witnessed low scores for suitability under COVID-19. Because of this, students who perceived family ($\mu=3.84$) and EI representatives ($\mu=3.69$): staff and students as their prime influencers in making their EI choice, have perceived greater suitability under COVID-19 than the students for whom prime influencers are preinstitute schools ($\mu=3.43$) and communities ($\mu=3.52$): relatives and friends.

Information sources ($F$-value=2.560, $p<0.05$) utilized by institutes in promoting their EIs do have a significant association with suitability under COVID-19. Interactive and face-to-face communications established by EI (before pandemic and after unlocking restrictions) through counselling programs and seminars at schools and online communication sources, EI websites, are evidenced to be of greater importance for students in making their institute choice during pandemics. This confirms that effective face-to-face communications such as counselling via

EI representatives, which are direct and reliable (Sia & Ming, 2010) (Yamamoto, 2006) and websites as a detailed source of valid information (Berge et al., 2019) (Briggs & Wilson, 2007), are important in making communication and spreading positive word-of-mouth about measures taken and hence suitability under COVID-19. Students receiving institutional information from social media, print materials, advertising media and outdoor activities have less bearing on suitability under COVID-19. Social media (M. L. Turner, 2017) (Le et al., 2019) a rich source of information and interactive communication in institute choice decisions, surprisingly, has a low impact on conveying the suitability of EI under COVID-19. The probable reason may be that the participants on social media are still not familiar with the present status and measures taken by the EI under the COVID-19 situation.

*5.3 Institute characteristics on suitability under pandemic*

The regression model exhibited moderate to strong accountability ($R$=0.659, $R^2$=0.435, $F$=35.220, $p$<0.000) in predicting the suitability of EI under COVID-19 by virtue of eleven characteristics affiliated with EIs. The statistical results revealed that EI characteristics such as proximity, image and reputation, quality education and curriculum delivery were the most significant contributors in predicting the suitability of EI under the COVID-19 pandemic (refer to Table IV). Other institute characteristics, such as location and locality, faculty profile, alumni profile, campus placements, infrastructure and facilities, safe and secured environment and value for money, are important characteristics in making institute choice insignificant predictors in estimating the suitability of EI under the COVID-19 pandemic.

This study has indicated that proximity (B=0.177, β=0.223, $t$=6.265, $p$<0.000) improves, i.e. more EI near the hometown, suitability under COVID-19 for EI increases. This is because EI's nearness to their hometown decreases the distance travelled, saves time and costs for the family (Chapman, 1981) and sustains health-related safety and security. More importantly, it reduces the risk of becoming inflected by coronavirus and creates appealing and promising conditions for EI selection (López Turley, 2009). This further justifies that EIs situated near students' markets are better in position to be selected by local students (Matusovich et al., 2020b), particularly in the pandemic situation.

This study revealed that the greater the perceived image and reputation (B=0.191, β=0.143, $t$=2.281, $p$<0.05) of EI, the greater the perception of the suitability of that EI under COVID-19. Word of mouth (Lafuente-Ruiz-de-Sabando et al., 2017) and trust and beliefs (Finch et al., 2013) are the key dimensions of image and reputation and are also an outcome of best practices followed by EIs to fulfill expectations over time on crisis management (Maringe & Gibbs, 2009). Furthermore, the buying behaviour of customers in a situational crisis is believed to be a function of organizational reputation and trust (Coombs, 1998). In summary, an EI with a good image and reputation is more likely to be trusted in providing suitable services and measures under the COVID-19 situation and increases students' desirability for institutions.

Quality education as a measure of institute choice (Joseph et al., 2005) has been cited as an overall excellence in service delivery (Sayeda et al., 2010). This study has revealed that quality education (B=0.363, β=0.268, $t$=4.017, $p$<0.05) is a significant predictor of suitability under COVID-19 and has a positive relationship that clarifies that a greater quality of education leads to a greater experience of suitability under the COVID-19 pandemic. Consequently, if the institute holds good-quality educational services, it is perceived to be right fit under COVID-19. A similar notion was made by (Zuhairi et al., 2020), who stated that quality education should remain at the forefront for effective learning during the COVID-19 pandemic situation..

Curriculum delivery during pandemics is the most difficult challenge for engineering studies and reshaping it in pandemics in an urgent need of the hour (Cahapay, 2020). In pandemic situations, successful curriculum delivery is entitled by gaining, accessing, and practicing knowledge, building skills and implementing the knowledge that keeps students' interest live through inculcating proper social distancing. This viewpoint is supported, as suitability under COVID-19 is significantly strengthened due to better curriculum delivery (B=0.186, $\beta$=0.141, $t$=2.073, $p$<0.05), referring to engineering majors. This study has identified it as a significant predictor for suitability under COVID-19.

Studies on online education with the use of ICT have shown that online learning has benefits in terms of its convenience, cost-effectiveness, and effort. (Zia, 2020a) documented that curriculum delivery is adversely poised out, as the curriculum was never designed by considering online delivery and why students are unable to cope with it during the pandemic. During pandemic infrastructure and facilities that are associated with computers/laptops, online learning platforms and cybersecurity are the more challenging issues for students (Neuwirth et al., 2020).

On the other hand, for (Alawamleh et al., 2020), onsite delivery is important to keep the academic concentration and motivation that gives true justice for competent engineers. This kind of digital and computing technology pattern is favourable for students who adopt computer allied majors, whereas it becomes unfavourable for students enrolled in non-computer allied majors who like machines and instruments to practice on. With this fact, students are more apprehensive about the nature of curriculum delivery that is better suited under the COVID-19 pandemic and keeps the value of engineering alive.

### *5.4 Institute characteristics on suitability under pandemic across students' characteristics*

Referring to Table VI, the following explanation is made for how institutional characteristics predicting suitability under COVID-19 differ across students' characteristics based on the regression model CS←(C1 to C11).

#### *5.4.1 Proximity*

Although proximity appeared to the bottom of the list of important characteristics of institutions in making EI choice decisions, the study showed that proximity (B=0.177, $\beta$=0.223, $t$=6.265, $p$<0.000) is an important characteristic in measuring suitability under COVID-19. As it improves, suitability under COVID-19 for that institute increases. In situations such as this pandemic, proximity to the hometown has become a prima facie for students from all backgrounds. Therefore, referring to this study, students segregated by gender, social class, native place, preinstitute performance, major choice, delivery mode, information source and human influence have all voices together to confirm that as it improves, the suitability of their EI under COVID-19 increases. This suggests that there is scope for local EIs in terms of their suitability under pandemic conditions for local students. In the case of STEM institutions, (Denzler & Wolter, 2011) found a similar advantage for institutions situated near local students.

#### *5.4.2 Location and locality*

During the COVID-19 pandemic, 'hot spots' or 'infected areas' related to coronavirus were the key anxieties for students; hence, they assessed it in terms of its spaciousness, airy ventilation, and health amenities. The importance of location and locality in EI choice is rated high (μ=4.120) by this study.

However, pertaining to its suitability under pandemic, for the group of high scorers (β= -0.18, $t$= -2.21, $p<0.05$), it turned out to be a possible risk of meeting virus infection by availing in such a facility and a source of wasting time that may indulge their study badly. Therefore, if the location and locality of the EI improves, the suitability of that EI under COVID-19 decreases.

Prominent computer allied majors are generally offered in EIs, which are located in a metropolitan area that is occupied by dense and highly populate localities. That is why students who joined computer allied majors think their urbanised location and locality of their EI is not suitable during the COVID-19 pandemic situation. Second, pedagogy related to computer allied majors is performed mostly in computer laboratories, which are supposed to have free access and are costly with air conditioning arrangements. If it is improved under pandemic conditions, they may be afraid of losing accessibility and comfort due to restrictions of social distancing norms. Therefore, for students opting computer allied majors (β= -0.16, $t$= -2.43, $p<0.05$), location and locality improve suitability under the COVID-19 pandemic. In contrast, non-computer allied majors are offered in every engineering EI surrounded by moderate location and locality. Accordingly, as per their beliefs, improved location and locality with required health facilities and amenities and onsite/hybrid pedagogy that is appropriate for learning non-allied majors by following social distancing norms are suitable for their engineering study under pandemic conditions. For this reason, students entering non-computer allied majors (β=0.26, $t$=2.92, $p<0.000$) have considered that better is the location and loyalty, better is the suitability under Covid-19. Similar findings were informed by (Joseph et al., 2005) about good location and locality, which is constructive in fetching enrolments into the campus, and (Ebell et al., 2020) mentioned it encouraging for in-person reporting of students.

Researchers (Berge et al., 2019) (Briggs & Wilson, 2007) have referred to websites as a detailed source of valuable information, an important source in making communication and spreading positive word-of-mouth that eventually attracts potential students. It is worth interpreting that because of website (β=0.31, $t$=2.98, $p<0.000$), these students were able to find the required information about location and locality which they found suitable under Covid-19 situation. Thus, students who perceived websites their prime source in selecting an EI of their choice perceived that the superior the location and locality were, the greater the suitability of that EI under COVID-19.

*5.4.3 Image and reputation*

Male students and students adopting reputed and high earning majors are noticed to be fascinated by EI carrying good image and reputation (MacLeod et al., 2017) (Munisamy et al., 2014) (Malgwi et al., 2005b). Due to this notion of image and reputation, this study also revealed that male students (β=0.20, $t$=2.60, $p<0.05$) and students enrolled in reputed computer allied majors (β=0.170, $t$=2.37, $p<0.05$) perceived that EI that having superior image and reputation provides greater suitability under COVID-19. (Gokuladas, 2010) identified similar findings where the computer allied profession is influenced by the organization's image and reputation.

In addition, image and reputation and higher social class are dignified with high esteem, high culture and high prestige, which becomes self-explanatory to why students belonging to higher social class ($β=0.21$, $t=2.78$, $p<0.05$) perceived greater suitability under COVID-19 with greater image and reputation. In contrast, (Davies & Guppy, 1997) confirmed that it is ineffective to fetch a group of lower social classes in engineering studies.

EI image and reputation is a function of tangible facilities, physical infrastructure, face-to-face interactions and support facilities (Plewa et al., 2016) provided by EI for effective curriculum delivery. This connoted that students who elected the onsite mode ($β=0.22$, $t=2.61$, $p<0.05$) during the pandemic perceived greater suitability of EI under COVID-19 because of their good image and reputation.

EI representatives such as students and staff are important information sources making positive word-of-mouth that boosts the image and reputation of institutions (Le et al., 2020). Students for whom EI staff and students ($β=0.19$, $t=2.20$, $p<0.05$) are influencing sources in making their institute choice have perceived greater suitability under the COVID-19 pandemic by virtue of the image and reputation of the institute. Furthermore, it makes clear that interactive face-to-face communications established by institutions (after unlocking restrictions) by means of counselling programs and seminars are a greater influencing source for students in clarifying the measures and suitability under COVID-19 that facilitate their institute choice. Other human influencers might not be familiar with the procedures and measures taken by the institute and hence fail to cogenerate communication and word-of-mouth that augment the image and reputation of EI under the COVID-19 situation.

*5.4.4 Faculty profile*

Faculty is a key influencer in building a strong relationship with students that affiliates a sense of belonging, confidence and satisfaction in students (Micari & Pazos, 2012). Faculty also act as facilitators and mentors in motivating, interacting and training students to achieve their career path (Mishkin et al., 2016) (Salami, 2007), the need of which is predominantly identified for underprivileged groups such as females and lower social classes under pandemic conditions. Next, these groups may be expecting faculty's expertise in terms of training assistance and support their emotions that they desperately require during the pandemic to improve their distress for better psychological well-being (Sood & Sharma, 2021). This creates greater realization about faculty members' profiles during the pandemic. Because of this, females ($β=0.28$, $t=2.59$, $p<0.05$) and students belonging to the lower social class ($β=0.26$, $t=2.22$, $p<0.05$) believed that institutions are more suitable under COVID-19 with a greater faculty profile. (Bao, 2020) documented similar conclusions in terms of the importance of faculty assistance, and the author found one of the five high-impacting pillars in sustaining higher education during the pandemic period. (Kolmos et al., 2013) reported similar results regarding the importance of mentors' roles in influencing female students.

Students with websites as a prime information source in their EI choice perceived suitability under pandemic less than expected due to the faculty profile. It can be concluded that faculty members' roles and responsibilities are clearly not notified in detail on websites during the pandemic. For this reason, students who adopted websites ($β= -0.27$, $t= -2.48$, $p<0.05$) as their information source experienced lower suitability for COVID-19 by virtue of the faculty profile.

*5.4.5 Alumni profile*

Alumni are another causative characteristic of institutions that are vital for potential students and their families in making their institute choice based on alumni's reputation gained after graduation (H.-F. Ho & Hung, 2008) and employment status (Kalimullin & Dobrotvorskaya, 2016). This appears to be true in this case, as students rated it highly important ($\mu=4.020$) in EI choice, yet it did not make any significant contribution to improving perceptions of suitability under COVID-19. This may be because alumni were not familiar with their EI's latest happenings during the pandemic because they remained inactive and inpersistent to impact potential students.

*5.4.6 Campus placements*

The majority of entry-level jobs in the engineering profession, particularly software and IT jobs, are offered mostly in industries or businesses that are situated in highly urbanized cities (Chen & Raveendran, 2012) accommodating highly compacted populace and are confirmed to be unsafe places from the COVID-19 point of view. During the pandemic, employment in industries and businesses located in these urbanized cities diminished. This is well thought by students residing in urban cities, as they are the main victims of it. This is why students whose hometown is an urban place ($\beta= -0.27$, $t= -2.48$, $p<0.05$) and students enrolled in computer allied majors ($\beta= -0.15$, $t= -1.83$, $p<0.1$) perceive that the greater the campus placement activities of EI are, the lower their perception of the suitability of EI under the COVID-19 pandemic.

*5.4.7 Quality education*

The institute's quality of education has significantly estimated the institute's suitability under COVID-19. Consequently, this reflects across students' groups, such as social class, native place, preinstitute performance and major choice, and is in a positive relationship that clarifies greater quality of education and greater experience with suitability under the COVID-19 pandemic. The applicability of quality education in regard to the competence, attitude, content, delivery and reliability of institute education (Sahney et al., 2004) by facilitating pedagogy to 'new normality' during the pandemic is essential.

The notion of 'quality' in higher education is best perceived and experienced due to tangible facilities, physical infrastructure and face-to-face human interactions and support that are sensed physically and not virtually. This nature of quality is thus better illustrated for onsite and hybrid delivery. Therefore, students considering hybrid and onsite learning are more concerned about quality infrastructure, facilities and services during the pandemic that keep their interest live and do not end in academic loss. Therefore, male students ($\beta=0.28$, $t=3.49$, $p<0.000$), students willing hybrid ($\beta=0.53$, $t=3.73$, $p<0.000$) and onsite ($\beta=0.33$, $t=3.40$, $p<0.000$) curriculum delivery are interested in competent physical and support facilities provided by the institute, which they think suitable under the COVID-19 pandemic. The importance of quality as discovered by (M. P. A. Murphy, 2020) (Yusuf & Jihan, 2020) for onsite arrangements made by EI by following social distancing norms is on the same grounds as observed in this study.

Students utilizing non-interactive communication ($\beta=0.48$, $t=2.90$, $p<0.05$), such as print material and media advertisements controlled by institutions and social media ($\beta=0.39$, $t=3.04$, $p<0.000$), similarly to human influence, such as family ($\beta=0.55$, $t=2.01$, $p<0.05$) and EI representatives ($\beta=0.27$, $t=3.02$, $p<0.000$), perceived that as quality education was enhanced, the suitability of COVID-19 increased. This means that these sources have created a better influence on students who cultivated the trust and reliability of EI under the COVID-19 pandemic on account of excellent quality provisions during the COVID-19 pandemic.

*5.4.8 Infrastructure and facilities*

It is a fundamental support system that needs to be rendered by an institute to continue learning during pandemic situations with safety and secured measures (Raaper & Brown, 2020). EI staff who are custodians and users of physical and tangible structures enrich their expertise in streamlining curricula for students. They, through their hands of experience on such an infrastructure and facilities, are better in a position to communicate its valuation to the students' market. Hence, students who joined EI due to the influence of EI staff ($\beta=0.20$, $t=2.53$, $p<0.05$) perceived that better infrastructure and facilities bring out superior suitability under COVID-19.

*5.4.9 Safe and secured environment*

Moreover, during pandemic situations, preventive measures after reopening EIs for students' overall wellbeing are the only visible way to survive the pandemic (Cheng et al., 2020). Today, provisions about such an ecosystem are contemplated to be mandatory standards and obligations in the reopening of EIs (UGC, 2021a). A safe and secured hygienic campus environment functions as a personal protection shield for students during pandemic situations. Urban students are well-experienced customers of these needs that maintain their health and safety intact. On this ground, it can be summarized that students belonging to urban hometowns have admitted superior suitability of EI under COVID-19 on behalf of their intrinsic need for safety, security and hygienic campus environments. (Gokuladas, 2010) revealed similar findings in which urban populations are most likely to be driven by intrinsic factors.

Improving safety and security measures for students with computer allied majors significantly increased suitability during the COVID-19 pandemic ($\beta=0.17$, $t=2.28$, $p<0.05$), as online pedagogy during the pandemic situation appeared to be safe and secured for them during the pandemic. Contradictory to this, students enrolled in non-computer allied majors perceived that with increasing safety and security measures, suitability of that EI under Covid-19 ($\beta= -0.22$, $t= -2.30$, $p<0.05$) reduces as they believe there can be an academic loss (Najmul Hasan & Bao, 2020) due to restricting the accessibility by over imposing safety and security.

*5.4.10 Curriculum delivery mode*

Curriculum delivery was found to be more common among male ($\beta=0.19$, $t=2.26$, $p<0.05$) students and high scorers ($\beta=0.35$, $t=3.29$, $p<0.05$), who believed that better suitability under the COVID-19 situation was due to better curriculum delivery. During the pandemic, with most of the education delivery being online, (Bisht et al., 2020) reported that Indian male students are discomforted with it. Recently, (Kundu & Bej, 2021) discovered in the Indian context that online learning during the pandemic is not suitable due to fear, insecurities, and a number of challenges associated with digital connections, (Wladis et al., 2014a) believed that online learning is concomitant with the risk of declining enrollments This might be the reason why curriculum delivery is crucial for male students during the pandemic. In contrast, (Kolmos et al., 2013) suggested that curriculum delivery should have minimum masculinity and be suitable for females for their inclusion in engineering.

Similarly, during the pandemic, the curriculum delivery of non-computer allied majors consisting of more numerical and practical-based subjects can trigger academic loss if properly not delivered. Second, higher working loads and skills gaps associated with these majors might be a worrying factor during the pandemic (Chadha et al., 2020). The hybrid delivery mode, a mixed version associated online for theory learning and onsite for practical and lab work, with safe distancing measures has shown positive signs (Tan, 2020) (Sia & Adamu, 2020) for students returning to campus after the reopening of EIs during the pandemic. This becomes a rational reason for the students of non-computer allied majors ($\beta=0.25$, $t=2.09$, $p<0.05$) and students' willingness to use the hybrid mode ($\beta=0.40$, $t=3.03$, $p<0.05$) to realize that superior suitability under COVID-19 is due to greater and sophisticated curriculum delivery. The findings of this study are in accordance with (Bordoloi, 2021), who uncovered that a hybrid version involving online and onsite sessions is the most suitable technique during the COVID-19 pandemic in the Indian context.

In terms of key sources of information and influence in making EI choices, students gathering information from face-to-face interactive counselling sessions ($\beta=0.49$, $t=2.79$, $p<0.05$) and students influenced by community (friends/peers, relatives) ($\beta=0.44$, $t=3.10$, $p<0.000$) perceived that better curriculum delivery moves suitability under COVID-19 in a positive direction. EI staff, as the eyewitness and practitioner, is the true communicator about disclosing details about curriculum delivery during the pandemic and hence governs the estimate of curriculum delivery on suitability under the pandemic.

*5.4.11 Value for money*

In the literature review, value for money for potential students is cited as rewarding benefits in terms of skill, quality, earnings, and social status against investment of parents' money, students' time, and efforts. During pandemic situations, cost-effectiveness, time and efforts spent are more vital, as they relate to mental and health conditions. For this cause of notion, in this study, value for money has some positive impressions. Studies pertaining to online education have shown that online learning is very cost-effective in terms of its conveniences and benefits (Alawamleh et al., 2020) (Tesar, 2020). For this reason, rural students ($\beta=0.14$, $t=2.03$, $p<0.05$), low performer students ($\beta=0.14$, $t=2.18$, $p<0.05$), students selecting online pedagogy ($\beta=0.29$, $t=2.75$, $p<0.05$) and students enrolled in computer-allied majors ($\beta=0.21$, $t=3.20$, $p<0.000$) may have emphasized greater suitability under COVID-19 for improved value for money. Referring to this study, it is proven to be appropriate.

However, in the case of students opting for hybrid pedagogy ($\beta=-0.27$, $t=-2.16$, $p<0.05$) and accessing both online and onsite pedagogy, the influence of value for money is negative. Hybrid delivery may result in doubling students' time, efforts, and cost, which incurred doe to blend pedagogy and therefore is not desired during pandemic situations. This might be the reason for these students to perceive that as value for money improves, suitability under COVID-19 decreases for them.

**6. Practical implications, visionary suggestions, and significance**

Based on the findings of this study, the following managerial implications and suggestions are envisioned for effective performance through the repositioning of EIs endorsed during the pandemic.

*6.1 Repositioning with best suitable practices during pandemic*

During the pandemic, proximity, image and reputation, quality of education and curriculum delivery are the prime governing characteristics for experiencing greater suitability under COVID-19, based on the results of the current study. Institutional practices associated with high esteem, high culture and high prestige are more significant in developing image and reputation. (Gill et al., 2018). EIs therefore should restructure their resources to achieve excellent governance. The image and reputation that coveys the quality of EIs (Hemsley-Brown & Oplatka, 2015a) can be earned by incorporating quality educational infrastructure and facilities along with effective risk-management measures during the pandemic. Later, mouth publicity was channelized through effective communications via online and face-to-face interactions (in small sizes) that furnished relationship building (Helgesen, 2008). With one action, it has two-fold benefits for EI during the pandemic. First, providing quality education and services will positively enhance image and reputation (Khoi et al., 2019). Second, it will build trust in EI's commitments to providing quality services and will achieve students' reliability and confidence in the quality provisions rendered by the EI during the pandemic. Institutions need to create indorse co-creating mechanisms to provide and process vital information about their offers to make informed choice decisions (Mogaji & Maringe, 2020).

Due to the immobility of institutions' physical assets, EI has little to work in proximity. However, this study has predicted the importance of proximity to the hometown in contributing to institutions' suitability for COVID-19 during the pandemic. It becomes binding on local institutes to provide excellent educational services with social distancing norms to grab this opportunity. Although institutes placed at a distance from their hometown can still accommodate good enrollments during pandemic situations by offering emerging and reputed majors, providing suitable delivery may be onsite or hybrid, as the situation permits them to keep the interest of students ongoing and minimize their academic loss. Rural and remote-based engineering campuses will have an advantage in providing onsite delivery, which can easily lead to academic and nonacademic pedagogy (Gross, 2020). However, to overcome proximity issues, such EIs must utilize effective promotional tools for cogenerating trust in EIs and must resilient to mitigation management, as framed by (UGC, 2021b). During the pandemic, if students could not make their way into the campus fully, the institute should reach them though virtually for some basic theoretical subjects. Then, bringing them for in-person pedagogy to deliver practical sessions or teaching difficult subjects in small groups on alternate days by following social distancing norms is the only visible solution for time being. However, such a hybrid delivery i.e. brick and click (Rana et al., 2020) should maintain balanced load by reshaping pedagogy (Cahapay, 2020) in terms of time consumed and efforts required. If the pandemic carries with us for a long life, then the institute will have to open up other options, in fact very few right now in the debate like: small campuses and relocating in remote places (Gross, 2020).

*6.2 Inclusion of diversity during pandemic*

Although the impact of suitability under pandemic is envisaged equally for the groups: gender, social class, native place, and pre-institute performance, remedies in making EI suitable should be analyses and executed at micro level, separate for each student coming out from different background to ensure inclusivity of diversity into the EI. The EI must realize that 'all sizes does not fit all'. The findings of the current study regarding the regression analysis of EI characteristics on suitability under pandemic conditions across student characteristics can be useful in this regard. For example, male students can be fueled into campus by providing quality concerns, suitable pedagogy delivery and stamping EI's image on their minds. This study has proven that these EI characteristics are promising in creating greater beliefs on the suitability of EI under pandemic conditions. A team of faculty members playing a role as counsellors or local guardians can be effective for underprivileged groups: female and lower social class students by creating interest and resilience (Nandy et al., 2021) that stimulate a sense of belonging during the pandemic that facilitates their institute choice. Higher social class and higher scorers who are strongly cognizant about image and reputation (Hemsley-Brown & Oplatka, 2015a) can be attracted by achieving rankings, publications in newspapers and gaining social status by participating in awareness campaigns about COVID-19 that directly will contribute to the image and reputation of the EI. Urban students coming from high cultured backgrounds should be provided with high-quality pedagogy facilities with safer and more secure amenities, particularly those related to health and dorm with hygienic ambience that are suitable during pandemics. EI's offers in terms of campus placements should be widened up their search towards remote and rural based industries / companies which are safest and secured place of making careers now a days as perceived by urban and computer allied enrollments of this study.

As per the findings of this study, despite the pandemic situation, students recommended onsite pedagogy and conceived suitable for their studies. This is true, as education is more than learning; it is much related to emotions, face reading and human touch, encouraging and channelizing students' social, affective and epistemic relations (Tarc, 2020), and further engineering education must have hands-on experience to be an ethical engineer. To keep the interest and concentration ongoing, students with high scores and non-computer allied majors must be exposed to suitable in-person delivery, hybrid or onsite depending on the intensity of the pandemic situation, and pedagogy can be delivered in small sizes under social distancing norms with all other amenities closed that consume time and occupy dense populate. Curriculum delivery to rural and low-scoring students should proceed in such a way that it should not create an extra burden of time, money, and effort and should not impose emotional stress and trauma due to technology handling. Likewise, online delivery should not incur extra charges for students due to the cost of digital devices and technology involved in the online mode.

EIs should inhouse all required facilities that meet diverse expectations of 'fit for the purpose' under the COVID-19 pandemic. The success of EIs will be dependent on how far and fast this 'fitness' creates a 'house of reliance' (Nandy et al., 2021) for the EI. This is where the role of human influence and EI's promotional tools are the keys to branding EI's suitability under COVID-19 in the market. EI stakeholders such as faculty, existing and alumni students who are real experience holders and direct sources of the EI are fit for the purpose of spreading 'word-of-mouth' and making its 'suitability' viral. However, if this pandemic becomes persistent in the future, then EIs to exhibit its existence must include an additional approach that presents their suitability through onsite and online existence. This includes making availability of EI information on stakeholders' personal digital devices on one click of their fingertips. This can be ensued by the detailed and sophisticated design of content, images and videos that facilitate stakeholders' interactions on EI online platforms such as websites and social media pages that produce a proper envision of EI attributes and quality concerns during pandemic crises. Alumni, current students, and faculty should be encouraged to be involved online and collaborative on LinkedIn and EI's YouTube channel, which facilitates the promotion of current happenings and live concerts on pedagogy during the pandemic. Other stakeholders, such as preinstitute schools and potential students with their parents, should be motivated to follow the EI Facebook page, which keeps them in touch with the EI. All such efforts will ultimately develop institute image (Manzoor et al., 2020) and quality of relationship (Clark et al., 2017), which is the need for hours foreseen in creating a future market for an engineering institute during a pandemic situation.

*6.3 Research contribution*

To the best of our knowledge, this study is the first to present insights into the performance of choice characteristics during the COVID-19 pandemic that are utilized to assess the selection of an engineering institute. Guessing is, it is also first to come up with new gadget 'suitability', which is noticed to be a significant utility in evaluating choice characteristics under pandemic conditions. It has successfully examined and explored the relationship of suitability and other traditional choice characteristics associated with EIs along with students' characteristics during the pandemic. This is the main contribution of this study. This may facilitate future research on creating best practices and guidelines for suitability.

The study has arrived with substantial hopes for academicians and policy makers that endorsed lucrative insights into how choice characteristics can be executed to regulate the 'suitability' of EI under the COVID-19 pandemic, which is in harmony with the inclusion of diversity. As it has firmly established and deeply rooted, EIs will benefit from administering 'new' enrollments, and aspiring students will be known about how to select 'new' EIs. While it is still a hot debate on how the COVID-19 pandemic will bring normality in EIs, this study has practically and statistically measured through a regression model that can be yardstick for staying ahead in a competitive market.

During this pandemic, enrollments this year had all the earmarks of being set for lost games. For academicians, it was a daily homework, and for marketing officers, it was repetitive guesswork. It is unclear how Indian EIs were able to withstand enrollments for the 2020-2021 academic year during this pandemic. However, even now, the findings of this study can be a game changer for them by understanding the characteristics governing the suitability of EI and repositioning themselves to provide the best suitable practices mentioned herein for holding and engaging engineering students. In the future, if the pandemic continues to be with us for a long time, this study is highly supportive of its revolutionary road, which is visible and feasible for bringing future students into EI campuses. Accordingly, the study has added new and substantial materials and thus has made several key contributions to the body of knowledge.

## 7. Conclusion

COVID-19 has impacted the higher education sector globally, including Indian EIs. It has tightened its knot around these EIs that enforced their previous half-shut shades completely down to prevent the risk of spreading COVID-19. EIs are now at more risk while doing nothing during the pandemic. Potential students live and grow for their life-dream 'college going', and this phenomenon does not occur in vacuum consequently for them. Regardless, the mindset of both EIs and potential students should be tailored to 'show must go on'. Fetching admissions to EI campuses before the pandemic was a difficult task, and during the pandemic, it became a challenge for survival. However, the current study has analytical mapped choice characteristics with suitability of EIs under pandemic conditions that are useful in choice decisions.

The main objective of this study was to examine choice characteristics and, consequently, to critically explore the relationships of EI characteristics with suitability during the COVID-19 pandemic based on student diversity. Three associated research questions were qualitatively answered, and three hypotheses were statistically validated. First, the study has noticed that conventional choice characteristics predominant in choice decisions are appropriate in pandemic situations. Second, major choice, curriculum delivery mode, human influence and information sources are the prejudiced characteristics of students that command the suitability of EIs under pandemic conditions. Next, the results confirmed that four EI characteristics, proximity, image and reputation, quality of education and curriculum delivery, contributed to the suitability of EIs. Specifically, at the micro level, the findings exposed multiple relationships between various EI characteristics and the suitability of EIs during the pandemic across student characteristics based on which EIs can reposition themselves to attract potentially diverse enrollments during the pandemic.

To culminate, at this moment, it is dubious that EIs will be weathering a 'new normality' during the pandemic. The answers to this question are very reliant on EI's resilience in meeting needs of diversity that integrates EI's characteristics instituted on student-centric suitability under pandemic conditions. Meanwhile, it is about an entire hearted endeavor that is needed during Covid-19 pandemic from all stakeholders may be in the form of influence, motivation, cooperation, coordination, training assistance, information sharing and much more intensifying demands for engineering education and facilitating choice decisions during pandemic period as a responsible citizen. This evolution may bring 'normality' to 'new' enrollments, can overwhelm lost decade, and can become revolutionary transformation in the future ahead!

## 8. Limitations and future research

Like any research that employs a limited sample, this study is restricted to the fact that it deals with a single context, North Maharashtra part of India, so that its findings cannot be directly generalized. All things considered, the current study's sincerity and relevance lies in exploring multilevel relationships of suitability (a new gadget) with traditional characteristics associated with institutions and students. Realizing these facts, plenty of research doors are open to investigating the impact of institutes offering other higher education in health-related studies, management, laws, or agriculture and institutions situated in different cultures. Such future studies may report various relationships, as choice characteristics vary with study majors and culture wherein the institute is situated; consequently, various perspectives on the suitability of the institute under pandemic conditions can be acquired.

Next, the survey was conducted during the COVID-19 pandemic, and the findings may not be similar in a normal situation. Another fact is that the choice process for students basically begins on their precollege school days. In India, as this pandemic arrived in 2020, some students might have responded to a few institute characteristics referring to its pre-pandemic situation (earlier to 2020). Henceforth, future research is encouraged periodically but frequently that includes the span of college choice representing the pandemic period. Last, the institute's characteristic of suitability under COVID-19 is utilized for the first time in this study to provide a general idea. Although sufficient progress on choice characteristics has been escorted in the first attempt, a more refined and detailed scale can be developed in future research.

## List of abbreviations

| | |
|---|---|
| AICTE: | All India Council for Technical Education, New Delhi |
| EI: | Engineering institution |
| HE: | Higher education |
| IT: | Information technology |
| ICT: | Information and communication technologies |
| UGC: | University grants commission, New Delhi |
| UNESCO: | United Nations Educational, Scientific and Cultural Organization |
| WHO: | World health organization |

# References

Abdullah, A. M., & Saeid, M. (2016). Factors affecting students' choice for MBA program in Kuwait Universities. *International Journal of Business and Management*, *11*(3), 119–128.

Abdullah, F. (2006). The development of HEdPERF: a new measuring instrument of service quality for the higher education sector. *International Journal of Consumer Studies*, *30*(6), 569–581.

Agarwala, T. (2008). Factors influencing career choice of management students in India. *Career Development International*, *13*(4), 362–376. https://doi.org/10.1108/13620430810880844

Ai, X., Chen, C., Xu, B., Zhang, M., Liu, Z., & Fu, G. (2018). A survey of college students' safety awareness. *Journal of Security and Safety Technology*, *6*(4), 84–91.

Aichouni, M., Benchicou, S., & Nehari, D. (2013). Knowledge management through the e-learning approach - a case study of online engineering courses. *European Journal of Engineering Education*, *38*(3), 316–328. https://doi.org/10.1080/03043797.2013.786025

AICTE New Delhi. (2021). *AICTE Dashboard*. http://www.aicte-india.org/dashboard/pages/graphs.php

AL-Mutairi, A., & Saeid, M. (2016). Factors Affecting Students' Choice for MBA Program in Kuwait Universities. *International Journal of Business and Management*, *11*(3), 119. https://doi.org/10.5539/ijbm.v11n3p119

Alawamleh, M., Al-Twait, L. M., & Al-Saht, G. R. (2020). The effect of online learning on communication between instructors and students during Covid-19 pandemic. *Asian Education and Development Studies*. https://doi.org/10.1108/AEDS-06-2020-0131

Alves, M., Rodrigues, C. S., Rocha, A. M. A. C., & Coutinho, C. (2016). Self-efficacy, mathematics' anxiety and perceived importance: an empirical study with Portuguese engineering students. *European Journal of Engineering Education*, *41*(1), 105–121. https://doi.org/10.1080/03043797.2015.1095159

Aristovnik, A., Keržič, D., Ravšelj, D., Tomaževič, N., & Umek, L. (2020). Impacts of the COVID-19 pandemic on life of higher education students: A global perspective. *Sustainability (Switzerland)*, *12*(20), 1–34. https://doi.org/10.3390/su12208438

Armstrong, J. J., & Lumsden, D. B. (2000). Impact of Universities' Promotional Materials on College Choice. *Journal of Marketing for Higher Education*, *9*(2), 83–91. https://doi.org/10.1300/J050v09n02

Ary, D., Jacobs, L. C., Sorensen, C. K., & Walker, D. A. (2010). *Introduction to Research in Education* (8th ed.). Wadsworth, Cengage Learning.

Aucejo, E. M., French, J., Araya, M. P. U., & Zafar, B. (2020). The impact of COVID-19 on student experiences and expectations: Evidence from a survey. *Journal of Public Economics*, *191*, 104271.

Ball, S. J., Davies, J., David, M., & Reay, D. (2002). "Classification'and'Judgement": social class and the'cognitive structures' of choice of Higher Education. *British Journal of Sociology of Education*, *23*(1), 51–72.

Bao, W. (2020). COVID -19 and online teaching in higher education: A case study of Peking University . *Human Behavior and Emerging Technologies*, *2*(2), 113–115. https://doi.org/10.1002/hbe2.191

Baytiyeh, H., & Naja, M. (2012). Identifying the challenging factors in the transition from colleges of engineering to employment. *European Journal of Engineering Education*, *37*(1), 3–14. https://doi.org/10.1080/03043797.2011.644761

Bearman, M., Smith, C. D., Carbone, A., Slade, S., Baik, C., Hughes-Warrington, M., & Neumann, D. L. (2012). Systematic review methodology in higher education. *Higher Education Research and Development*, *31*(5), 625–640. https://doi.org/10.1080/07294360.2012.702735

Bennet, D. (2006). The effectiveness of current student ambassadors in HE marketing recruitment and retention. *International Conference on HE Marketing Cyprus*, 3–5.

Berge, M., Silfver, E., & Danielsson, A. (2019). In search of the new engineer: gender, age, and social class in information about engineering education. *European Journal of Engineering Education*, *44*(5), 650–665. https://doi.org/10.1080/03043797.2018.1523133

Berjano, E., Sales-Nebot, L., & Lozano-Nieto, A. (2013). Improving professionalism in the engineering curriculum through a novel use of oral presentations. *European Journal of Engineering Education*, *38*(2), 121–130. https://doi.org/10.1080/03043797.2012.745829

Bernard, R. M., Abrami, P. C., Lou, Y., Borokhovski, E., Wade, A., Wozney, L., Wallet, P. A., Fiset, M., & Huang, B. (2004). How does distance education compare with classroom instruction? A meta-analysis of the empirical literature. *Review of Educational Research*, *74*(3), 379–439.

Bhattacharya, B. (2004). What is "good teaching" in engineering education in India? A case study. *Innovations in Education and Teaching International*, *41*(3), 329–341. https://doi.org/10.1080/14703290410001733258

Bird, K. A. (2020). *Negative impacts from the shift to online learning during the COVID-19 crisis: Evidence from a statewide community college system*. 20.

Bird, K., Castleman, B. L., & Lohner, G. (2020). *Negative impacts from the shift to online learning during the COVID-19 crisis: evidence from a statewide community college system*.

Bisht, R. K., Jasola, S., & Bisht, I. P. (2020). Acceptability and challenges of online higher education in the era of COVID-19: a study of students' perspective. *Asian Education and Development Studies*. https://doi.org/10.1108/AEDS-05-2020-0119

Bitner, M. J. (1992). Using Background Music to Affect the Behaviour of Supermarket Shoppers. *Journal of Marketing*, *56*(2), 57. https://doi.org/10.2307/1252042

Blom, A., & Cheong, J. (2010). *Governance of Technical Education in India: Key Issues, Principles, and Case Studies.* (A. Blom & J. Cheong (eds.)). The World Bank. https://play.google.com/books/reader?id=Vc8QCaV_V9sC&pg=GBS.PA16

Blom, Andreas, & Saeki, H. (2011). *Employability and Skill Set of Newly Graduated Engineers in India.* https://doi.org/10.1596/1813-9450-5640

Bodycott, P. (2009). Choosing a higher education study abroad destination: What mainland Chinese parents and students rate as important. *Journal of Research in International Education*, *8*(3), 349–373.

Bordoloi, R. (2021). *Perception towards online / blended learning at the time of Covid-19 pandemic : an academic analytics in the Indian context.* https://doi.org/10.1108/AAOUJ-09-2020-0079

Bordón, P., Canals, C., & Mizala, A. (2020). The gender gap in college major choice in Chile. *Economics of Education Review*, *77*(June), 102011. https://doi.org/10.1016/j.econedurev.2020.102011

Borrego, M., Douglas, E., & Amelink, C. (2009). Quantitative, qualitative, and mixed research methods in engineering education. *Journal of Engineering Education*, *98*(1), 55–66. https://doi.org/10.1002/j.2168-9830.2009.tb01005.x

Briggs, S. (2006). An exploratory study of the factors influencing undergraduate student choice: The case of higher education in Scotland. *Studies in Higher Education*, *31*(6), 705–722. https://doi.org/10.1080/03075070601004333

Briggs, S. R., & Cheek, J. M. (1986). The role of factor analysis in the development and evaluation of personality scales. *Journal of Personality*, *54*(1), 106–148. https://doi.org/10.1111/j.1467-6494.1986.tb00391.x

Briggs, S., & Wilson, A. (2007a). Which university? A study of the influence of cost and information factors on Scottish undergraduate choice. *Journal of Higher Education Policy and Management*, *29*(1), 57–72.

Briggs, S., & Wilson, A. (2007b). Which university? A study of the influence of cost and information factors on Scottish undergraduate choice. *Journal of Higher Education Policy and Management*, *29*(1), 57–72. https://doi.org/10.1080/13600800601175789

Broekemier, G. M., & Seshadri, S. (2000). Differences in college choice criteria between deciding students and their parents. *Journal of Marketing for Higher Education*, *9*(3), 1–13.

Brown, P. R., McCord, R. E., Matusovich, H. M., & Kajfez, R. L. (2015). The use of motivation theory in engineering education research: a systematic review of literature. *European Journal of Engineering Education*, *40*(2), 186–205. https://doi.org/10.1080/03043797.2014.941339

Bruce, G., & Edgington, R. (2008). Factors Influencing Word-of-Mouth Recommendations by MBA Students: An Examination of School Quality, Educational Outcomes, and Value of the MBA. *Journal of Marketing for Higher Education*, *16*(1), 79–101. https://doi.org/10.1080/08841240802100303

Cahapay, M. B. (2020). Rethinking Education in the New Normal Post-COVID-19 Era: A Curriculum Studies Perspective. *Aquademia*, *4*(2), ep20018. https://doi.org/10.29333/aquademia/8315

Calitz, A. P., Cullen, M. D. M., & Jooste, C. (2020). The Influence of Safety and Security on Students' Choice of University in South Africa. *Journal of Studies in International Education*, *24*(2), 269–285. https://doi.org/10.1177/1028315319865395

Caroni, C. (2011). Graduation and attrition of engineering students in Greece. *European Journal of Engineering Education*, *36*(1), 63–74. https://doi.org/10.1080/03043797.2010.539676

Case, J. M., Fraser, D. M., Kumar, A., & Itika, A. (2016). The significance of context for curriculum development in engineering education: a case study across three African countries. *European Journal of Engineering Education*, *41*(3), 279–292. https://doi.org/10.1080/03043797.2015.1056103

Cattaneo, M., Horta, H., Malighetti, P., Meoli, M., & Paleari, S. (2017). Effects of the financial crisis on university choice by gender. *Higher Education*, *74*(5), 775–798. https://doi.org/10.1007/s10734-016-0076-y

CDC. (2021). *Considerations for institutions of higher education*. Retrieved Jan. https://www.cdc.gov/coronavirus/2019-ncov/community/colleges-universities/considerations.html?CDC_AA_refVal=https%3A%2F%2Fwww.cdc.gov%2Fcoronavirus%2F2019-ncov%2Fcommunity%2Fguidance-ihe-response.html

Cebr. (2016). *Engineering and economic growth: a global view* (Issue September). http://www.raeng.org.uk/publications/reports/engineering-and-economic-growth-a-global-view

Chadha, D., Kogelbauer, A., Campbell, J., Hellgardt, K., Maraj, M., Shah, U., Brechtelsbauer, C., & Hale, C. (2020). Are the kids alright? Exploring students' experiences of support mechanisms to enhance wellbeing on an engineering programme in the UK. *European Journal of Engineering Education*, *0*(0), 1–16. https://doi.org/10.1080/03043797.2020.1835828

Chakrabarti, A. (2014). Determinants of Participation in Higher Education and Choice of Disciplines : Evidence from Urban and Rural Indian Youth. *South Asia Economic Journal*, *2*(2009), 371–402. https://doi.org/10.1177/139156140901000205

Chanana, K. (2000). Treading the Hallowed Halls: Women in Higher Education in India. *Economic and Political Weekly*, *35*(12), 1012–1022. http://www.jstor.org/stable/4409055

Chapman, D. (1981). A Model of Student College Choice. *The Journal of Higher Education*, *52*(5), 490–505. https://doi.org/10.2307/1981837

Chapman, D. W., & Johnson, R. H. (1979). Influences on students' college choice: A case study. *Ann Arbor, MI: Project CHOICE, School of Education, University of Michigan*.

Chattefuee, S., & Hadi, A. S. (2006). Regression Analysis by Example: Fourth Edition. In *Regression Analysis by Example: Fourth Edition*. https://doi.org/10.1002/0470055464

Chauhan, K., & Pillai, A. (2013). Role of content strategy in social media brand communities: a case of higher education institutes in India. *Journal of Product & Brand Management*, *22*(1), 40–51. https://doi.org/10.1108/10610421311298687

Chen, M. A., & Raveendran, G. (2012). Urban Employment in India: Recent Trends and Patterns. *Margin*, *6*(2), 159–179. https://doi.org/10.1177/097380101200600204

Cheng, S. Y., Wang, C. J., Shen, A. C. T., & Chang, S. C. (2020). How to Safely Reopen Colleges and Universities During COVID-19: Experiences From Taiwan. *Annals of Internal Medicine*, *173*(8), 638–641. https://doi.org/10.7326/M20-2927

Chin, W. W. (1998). The partial least squares approach to structural equation modelling. In Marcoulides G. A. (Ed.). *Modern Methods for Business Research*, *295*(2), 295–336.

Chugh, R., & Ruhi, U. (2018). Social media in higher education: A literature review of Facebook. *Education and Information Technologies*, *23*(2), 605–616. https://doi.org/10.1007/s10639-017-9621-2

Churchill Jr, G. A. (1979). A Paradigm for Developing Better Measures of Marketing Constructs. *Journal of Marketing Research*, *XVI*(February), 64–73.

Ciotti, M., Ciccozzi, M., Terrinoni, A., Jiang, W. C., Wang, C. Bin, & Bernardini, S. (2020). The COVID-19 pandemic. *Critical Reviews in Clinical Laboratory Sciences*, *57*(6), 1–24. https://doi.org/10.1080/10408363.2020.1783198

Clark, M., Fine, M. B., & Scheuer, C.-L. (2017). Relationship quality in higher education marketing: the role of social media engagement. *Journal of Marketing for Higher Education*, *27*(1), 40–58. https://doi.org/10.1080/08841241.2016.1269036

Cohen, J. (1988). *Statistical power analysis for the behavioural sciences. Hillsdale, NJ: Laurence Erlbaum Associates*. Inc.

Cohen, J., Cohen, P., West, S. G., & Aiken, L. S. (2013). Applied Multiple Regression/Correlation Analysis for the Behavioral Sciences. *Applied Multiple Regression/Correlation Analysis for the Behavioral Sciences*. https://doi.org/10.4324/9780203774441

Cohen, L., Manion, L., & Morrison, K. (2007). Research Methods in Education. In *British Journal of Educational Studies* (Vol. 55, Issue 3). https://doi.org/10.1111/j.1467-8527.2007.00388_4.x

Conard, M. J., & Conard, M. A. (2000). An analysis of academic reputation as perceived by consumers of higher education. *Journal of Marketing for Higher Education*, *9*(4), 69–80. https://doi.org/10.1300/J050v09n04_05

Coombs, W. T. (1998). An Analytic Framework for Crisis Situations: Better Responses From a Better Understanding of the Situation. *Journal of Public Relations Research*, *10*(3), 177–191. https://doi.org/10.1207/s1532754xjprr1003_02

Creswell, J. W. (2012a). Educational research: Planning, conducting, and evaluating quantitative and qualitative research. In *Educational Research* (Vol. 4). https://doi.org/10.1017/CBO9781107415324.004

Creswell, J. W. (2012b). The Process of Conducting Research Using Quantitative and Qualitative Approaches. *Educational Research: Planning, Conducting, and Evaluating Quantitative and Qualitative Research*, 1–25. https://doi.org/10.1080/08832323.1963.10116709

Creswell, J. W., & Creswell, J. D. (2017). *Research design: Qualitative, quantitative, and mixed methods approaches*. Sage publications.

Cronbach, L. J. (1951). Coefficient alpha and the internal structure of tests. *Psychometrika*, *16*(3), 297–334.

Daoud, J. I. (2018). Multicollinearity and Regression Analysis. *Journal of Physics: Conference Series*, *949*(1). https://doi.org/10.1088/1742-6596/949/1/012009

Davies, S., & Guppy, N. (1997). Fields of study, college selectivity, and student inequalities in higher education. *Social Forces*, *75*(4), 1417–1438. https://doi.org/10.1093/sf/75.4.1417

Dawes, P. L., & Brown, J. (2002). Determinants of Awareness, Consideration, and Choice Set Size in University Choice. *Journal of Marketing for Higher Education*, *12*(1), 49–75. https://doi.org/10.1300/J050v12n01

De Courcy, J. W. (1987). The Quality of Realisation. *European Journal of Engineering Education*, *12*(3), 213–218. https://doi.org/10.1080/03043798708939363

Denzler, S., & Wolter, S. (2011). Too Far to Go? Does Distance Determine Study Choices? In *IZA Discussion Paper* (No. 5712). http://nbn-resolving.de/urn:nbn:de:101:1-201105306388

DesJardins, S. L., & Toutkoushian, R. K. (2005). Are students really rational? The development of rational thought and its application to student choice. In *Higher education: Handbook of theory and research* (pp. 191–240). Springer.

Downey, G. L., & Lucena, J. C. (2005). National identities in multinational worlds: engineers and'engineering cultures'. *International Journal of Continuing Engineering Education and Life Long Learning*, *15*(3–6), 252–260.

Ebell, M. H., Bierema, L., & Haines, L. (2020). Graduate students significantly more concerned than undergraduates about returning to campus in the era of COVID-19. *MedRxiv*, 1–11. https://doi.org/10.1101/2020.07.15.20154682

Elliott, K. M., & Healy, M. a. (2001). Key Factors Influencing Student Satisfaction Related to Recruitment and Retention Key Factors Influencing Student Satisfaction Related to Recruitment and Retention. *Journal of Marketing for Higher Education*, *10*(4), 1–11. https://doi.org/10.1300/J050v10n04

Etikan, I., Musa, S. A., & Alkassim, R. S. (2016). Comparison of convenience sampling and purposive sampling. *American Journal of Theoretical and Applied Statistics*, *5*(1), 1–4.

Evans, C. A., Chen, R., & Hudes, R. P. (2020). Understanding Determinants for STEM Major Choice Among Students Beginning Community College. *Community College Review*, *48*(3), 227–251. https://doi.org/10.1177/0091552120917214

Ferry, T. R., Fouad, N. A., & Smith, P. L. (2000). The Role of Family Context in a Social Cognitive Model for Career-Related Choice Behavior: A Math and Science Perspective. *Journal of Vocational Behavior*, *57*(3), 348–364. https://doi.org/10.1006/jvbe.1999.1743

Finch, D., McDonald, S., & Staple, J. (2013). Reputational interdependence: an examination of category reputation in higher education. *Journal of Marketing for Higher Education*, *23*(1), 34–61. https://doi.org/10.1080/08841241.2013.810184

Floyd, F., & Widaman, F. (1995). Factor Analysis in the Development and Refinement of Clinical Assessment Instruments. *Psychological Assessment*, *7*(3), 286–299. https://doi.org/10.1037/1040-3590.7.3.286

Foskett, N., & Hemsley-Brown, J. (2001). *Choosing futures: Young people's decision-making in education, training, and careers markets*. Psychology Press.

Fox, J. (2016). *Applied Regression Analysis and Gereralized Linear Models*.

Freeman, K. (2012). *African Americans and college choice: The influence of family and school*. SUNY Press.

Freeman, K., & Brown, M. C. (2005). African Americans and college choice: The influence of family and school. *African Americans and College Choice: The Influence of Family and School*, 1–132.

Frenette, M. (2006). Too far to go on? Distance to school and university participation. In *Education Economics* (Vol. 14, Issue 1). https://doi.org/10.1080/09645290500481865

Gajić, J. (2012a). Importance of Marketing Mix in Higher Education Institutions. *Značaj Marketing Miksa U Visokoobrazovnim Institucijama.*, *9*(1), 29–41. https://doi.org/10.5937/sjas1201029G

Gajić, J. (2012b). Importance of Marketing Mix in Higher Education Institutions. *Singidunum Journal*, *9*(1), 29–41.

Gajić, J., Živković, R., & Stanić, N. (2017). KEY ATTRIBUTES OF SUCCESSFUL COMMUNICATION BETWEEN HIGHER EDUCATION INSTITUTION AND PROSPECTIVE STUDENTS. *TEME: Casopis Za Društvene Nauke*, *41*(3).

Gambhir, V., Wadhwa, N. C., & Grover, S. (2013). Interpretive structural modelling of enablers of quality technical education: an Indian perspective. *International Journal of Productivity and Quality Management*, *12*(4), 393. https://doi.org/10.1504/IJPQM.2013.056734

Gautam, D. K., & Gautam, P. K. (2020). Transition to Online Higher Education during COVID-19 Pandemic: Turmoil and Way Forward to Developing Country-Nepal. *Research Square*. https://doi.org/10.1108/JRIT-10-2020-0051

Gautam, M. (2015). Gender, Subject Choice and Higher Education in India. *Contemporary Education Dialogue*, *12*(1), 31–58. https://doi.org/10.1177/0973184914556865

Gibbs, P., & Knapp, M. (2012). *Marketing higher and further education: An educator's guide to promoting courses, departments and institutions* (Vol. 9781136609). Routledge.

Gill, T., Vidal Rodeiro, C., & Zanini, N. (2018). Higher education choices of secondary school graduates with a Science, Technology, Engineering or Mathematics (STEM) background. *Journal of Further and Higher Education*, *42*(7), 998–1014. https://doi.org/10.1080/0309877X.2017.1332358

Glenn, N. D., & Shelton, B. A. (1983). Pre-adult background variables and divorce: A note of caution about overreliance on explained variance. *Journal of Marriage and the Family*.

Godwin, A., Potvin, G., Hazari, Z., & Lock, R. (2016a). Identity, critical agency, and engineering: An affective model for predicting engineering as a career choice. *Journal of Engineering Education*, *105*(2), 312–340.

Godwin, A., Potvin, G., Hazari, Z., & Lock, R. (2016b). Identity, Critical Agency, and Engineering: An Affective Model for Predicting Engineering as a Career Choice. *Journal of Engineering Education*, *105*(2), 312–340. https://doi.org/10.1002/jee.20118

Goff, B., Patino, V., & Jackson, G. (2004). Preferred information sources of high school

students for community colleges and universities. *Community College Journal of Research and Practice*, *28*(10), 795–803. https://doi.org/10.1080/10668920390276957

Goi, C., & Ng, P. Y. (2008). E-learning in Malaysia: Success factors in implementing e-learning program. *International Journal of Teaching and Learning in Higher Education*, *20*(2).

Gokuladas, V. K. (2010). Factors that influence first-career choice of undergraduate engineers in software services companies. *Career Development International*, *15*(2), 144–165. https://doi.org/10.1108/13620431011040941

Gray, M., & Daugherty, M. (2004). Factors that Influence Students to Enroll in Technology Education Programs. *The Journal of Technology Education*, *15*(2), 5–19. https://doi.org/10.21061/jte.v15i2.a.1

Griffith, A. L., & Rothstein, D. S. (2009). Can't get there from here: The decision to apply to a selective college. *Economics of Education Review*, *28*(5), 620–628.

Gross, K. (2020). For Some Small Colleges, The Pandemic Could Sadly Be Their Savior. *New England Journal of Higher Education*.

Gupta, N. (2012). Women Undergraduates in Engineering Education in India: A Study of Growing Participation. *Gender, Technology and Development*, *16*(2), 153–176. https://doi.org/10.1177/097185241201600202

Gurukkal, R. (2020). Will COVID 19 Turn Higher Education into Another Mode? *Higher Education for the Future*, *7*(2), 89–96. https://doi.org/10.1177/2347631120931606

Han, P. (2014). A Literature Review on College Choice and Marketing Strategies for Recruitment. *Family and Consumer Sciences Research Journal*, *43*(2), 120–130. https://doi.org/10.1111/fcsr.12091

Hannagan, T. (1992). *Marketing for the non-profit sector*. Springer.

Hardy, D. E., & Katsinas, S. G. (2007). Classifying community colleges: How rural community colleges fit. *New Directions for Community Colleges*, *2007*(137), 5–17.

Hasan, Najmul, & Bao, Y. (2020). Impact of "e-Learning crack-up" perception on psychological distress among college students during COVID-19 pandemic: A mediating role of "fear of academic year loss." *Children and Youth Services Review*, *118*, 105355.

Hasan, Naziya, & Khan, N. H. (2020). *ONLINE TEACHING-LEARNING DURING COVID-19 PANDEMIC : STUDENTS '*. October.

Hearn, J. C. (1988). Attendance at higher-cost colleges: Ascribed, socioeconomic, and academic influences on student enrollment patterns. *Economics of Education Review*, *7*(1), 65–76. https://doi.org/10.1016/0272-7757(88)90072-6

Heathcote, D., Savage, S., & Hosseinian-Far, A. (2020). Factors Affecting University Choice Behaviour in the UK Higher Education. *Education Sciences*, *10*(8), 199.

Helgesen, Ø. (2008). Marketing for higher education: A relationship marketing approach. *Journal of Marketing for Higher Education*, *18*(1), 50–78. https://doi.org/10.1080/08841240802100188

Hemmo, V., & Love, P. (2008). *Encouraging student interest in science and technology*

*studies*. OECD Publishing.

Hemsley-Brown, J., & Oplatka, I. (2015a). *Higher education consumer choice*. Springer.

Hemsley-Brown, J., & Oplatka, I. (2015b). University choice: what do we know, what don't we know and what do we still need to find out? *International Journal of Educational Management*. https://doi.org/10.1108/IJEM-10-2013-0150

Hemsley-Brown, J., & Oplatka, I. (2015c). University choice: What do we know, what don't we know and what do we still need to find out? *International Journal of Educational Management*, *29*(3), 254–274. https://doi.org/10.1108/IJEM-10-2013-0150

Hicks, T., & Wood, J. L. (2016). A meta-synthesis of academic and social characteristic studies: First-generation college students in STEM disciplines at HBCUs. *Journal for Multicultural Education*, *10*(2), 107–123. https://doi.org/10.1108/JME-01-2016-0018

Hnatkovska, V., & Lahiri, A. (2013). The Rural-Urban Divide in India. *International Growth Centre Working Paper*, *February*, 1–24. http://www.theigc.org/wp-content/uploads/2014/09/Hnatkovska-Lahiri-2012-Working-Paper-March.pdf

Ho, G. K. S., & Law, R. (2020). Marketing Strategies in the Decision-Making Process for Undergraduate Choice in Pursuit of Hospitality and Tourism Higher Education: The Case of Hong Kong. *Journal of Hospitality and Tourism Education*, *00*(00), 1–13. https://doi.org/10.1080/10963758.2020.1791136

Ho, H.-F., & Hung, C.-C. (2008). Marketing mix formulation for higher education: An integrated analysis employing analytic hierarchy process, cluster analysis and correspondence analysis. *International Journal of Educational Management*, *22*(4), 328–340. https://doi.org/10.1108/09513540810875662

Ho, H., & Hung, C. (2008). Marketing mix formulation for higher education: An integrated analysis employing analytical hierarchy process, cluster analysis and correspondence analysis. *International Journal of Educational Management*, *22*(4), 328–340. https://doi.org/10.1108/09513540810875662

Holdsworth, D. K., & Nind, D. (2006). Choice modeling New Zealand high school seniors' preferences for university education. *Journal of Marketing for Higher Education*, *15*(2), 81–102.

Horstschräer, J. (2012). University rankings in action? The importance of rankings and an excellence competition for university choice of high-ability students. *Economics of Education Review*, *31*(6), 1162–1176.

Hossler, D., Braxton, J., & Coopersmith, G. (1989a). Understanding student college choice. *Higher Education: Handbook of Theory and Research*, *5*, 231–288.

Hossler, D., Braxton, J., & Coopersmith, G. (1989b). Understanding student college choice. In *Higher education: Handbook of theory and research* (Vol. 5, Issue January 1989). Agathon press New york.

Huntington-Klein, N. (2018). College Choice As a Collective Decision. *Economic Inquiry*, *56*(2), 1202–1219. https://doi.org/10.1111/ecin.12470

IAU. (2020). *The Impact of Covid-19 on Higher Education around the World*. https://www.iau-aiu.net/IMG/pdf/iau_covid19_and_he_survey_report_final_may_2020.pdf

Iloh, C. (2019). An Alternative to College" Choice" Models and Frameworks: The Iloh Model of College-Going Decisions and Trajectories. *College and University*, *94*(4), 2–9.

Imenda, S. N., Kongolo, M., & Grewal, A. S. (2004). Factors Underlying Technikon and University Enrolment Trends in South Africa. *Educational Management Administration & Leadership*, *32*(2), 195–215. https://doi.org/10.1177/1741143204041884

Ivy, J. (2002). University image: the role of marketing in MBA student recruitment in state subsidised universities in the Republic of South Africa. *Unpublished Doctoral Dissertation, University of Leicester, Leicester*.

Ivy, Jonathan. (2008). A new higher education marketing mix: the 7Ps for MBA marketing. *International Journal of Educational Management*, *22*(4), 288–299. https://doi.org/10.1108/09513540810875635

Izquierdo, F. A. (1993). Quality-designed curricula. *European Journal of Engineering Education*, *18*(4), 339–344. https://doi.org/10.1080/03043799308923253

Jacobs, J. A. (1986). The sex-segregation of fields of study: Trends during the college years. *The Journal of Higher Education*, *57*(2), 134–154.

Jain, R., Sahney, S., & Sinha, G. (2013). Developing a scale to measure students' perception of service quality in the Indian context. *TQM Journal*, *25*(3), 276–294. https://doi.org/10.1108/17542731311307456

Jarvie-Eggart, M. E., Singer, A. M., & Mathews, J. (2020). *Parent and family influence on first-year engineering major choice*.

Jayalath, C., Wickramasinghe, U., & Kottage, H. (2020). Factors Influencing Orderly Transition to Online Deliveries during COVID19 Pandemic Impact. *Asian Journal of Education and Social Studies*, *9*(2), 10–24. https://doi.org/10.9734/AJESS/2020/v9i230242

Jeffries, D., Curtis, D. D., & Conner, L. N. (2020). Student Factors Influencing STEM Subject Choice in Year 12: a Structural Equation Model Using PISA/LSAY Data. *International Journal of Science and Mathematics Education*, *18*(3), 441–461. https://doi.org/10.1007/s10763-019-09972-5

Jha, S. (2017). *Socio-economic Determinants of Inter-state Student Mobility in India : Implications for Higher Education Policy*. *27*. https://doi.org/10.1177/2347631117708069

Joseph, M., Yakhou, M., & Stone, G. (2005). An educational institution's quest for service quality: Customers' perspective. *Quality Assurance in Education*, *13*(1), 66–82. https://doi.org/10.1108/09684880510578669

Kalimullin, A. M., & Dobrotvorskaya, S. G. (2016). Higher Education Marketing Strategies Based on Factors Impacting the Enrollees' Choice of a University and an Academic Program. *International Journal of Environmental and Science Education*, *11*(13), 6025–6040.

Kallio, R. E. (1995). Factors influencing the college choice decisions of graduate students. *Research in Higher Education*, *36*(1), 109–124. https://doi.org/10.1007/BF02207769

Karataş, F. Ö., Bodner, G. M., & Unal, S. (2016). First-year engineering students' views of the nature of engineering: implications for engineering programmes. *European Journal*

*of Engineering Education*, *41*(1), 1–22.

Kelly, R., McGarr, O., Lehane, L., & Erduran, S. (2019). STEM and gender at university: focusing on Irish undergraduate female students' perceptions. *Journal of Applied Research in Higher Education*, *11*(4), 770–787. https://doi.org/10.1108/JARHE-07-2018-0127

Kenneth, N. R. (2005). Quantitative Research Methods in Educational Planning. In *Quantitative Research Methods in Educational Planning*. UNESCO International Institute for Educational Planning.

Kern, C. W. K. (2000). College choice influences: Urban high school students respond. *Community College Journal of Research & Practice*, *24*(6), 487–494.

Khanna, M., Jacob, I., & Yadav, N. (2014). Identifying and analyzing touchpoints for building a higher education brand. *Journal of Marketing for Higher Education*, *24*(1), 122–143. https://doi.org/10.1080/08841241.2014.920460

Khoi, B. H., Dai, D. N., Lam, N. H., & Chuong, N. Van. (2019). The relationship among education service quality, university reputation and behavioral intention in vietnam. In *Studies in Computational Intelligence* (Vol. 809). Springer International Publishing. https://doi.org/10.1007/978-3-030-04200-4_21

Kim, H. K., & Niederdeppe, J. (2013). The role of emotional response during an H1N1 influenza pandemic on a college campus. *Journal of Public Relations Research*, *25*(1), 30–50.

Kinnunen, P., Butler, M., Morgan, M., Nylen, A., Peters, K., Sinclair, J., Kalvala, S., Pesonen, E., Kinnunen, P., Butler, M., Morgan, M., & Nylen, A. (2016). *Understanding initial undergraduate expectations and identity in computing studies*. *3797*(February). https://doi.org/10.1080/03043797.2016.1146233

Kolb, B. (2008). Marketing Research for and Creative Organizations. In *Marketing Research*.

Kolmos, A., Mejlgaard, N., Haase, S., & Holgaard, J. E. (2013). Motivational factors, gender and engineering education. *European Journal of Engineering Education*, *38*(3), 340–358. https://doi.org/10.1080/03043797.2013.794198

Kotler, P, & Fox, K. F. M. (1995). *Strategic marketing for educational institutions*. *1546*(July).

Kotler, Philip, & Armstrong, G. (2013). *Principles of Marketing (16th Global Edition)*. Harlow: Pearson.

Kotler, Philip, Armstrong, G., Saunders, J., & Wong, V. (2016). *Principles of Marketing* (4th ed.). Prentice Hall Europe.

Kotler, Philip, & Fox, K. (1985). Strategic Marketing for Educational Institutions Prentice Hall. *Engelwood Cliffs, NJ*.

Kotler, Philip, & Fox, K. F. A. (1995). *Strategic marketing for educational institutions*. Prentice Hall.

Kotler, Philip, Hayes, T., & Bloom, P. N. (2002). *Marketing professional services*. Prentice Hall.

Kotler, Philip, Keller, K. L., Brady, M., & Goodman, M. (2009). Marketing Management. In

*Pearson Education Limited*. Pearson Education Limited. https://books.google.co.za/books?id=6uU-Dz-sCIQC

Kowalik, E. (2011). Engaging alumni and prospective students through social media. In *Cutting-Edge Technologies in Higher Education* (Vol. 2). Emerald. https://doi.org/10.1108/S2044-9968(2011)0000002014

Kundu, A., & Bej, T. (2021). COVID-19 response: students' readiness for shifting classes online. *Corporate Governance: The International Journal of Business in Society*.

Kutnick, P., Chan, R. Y. Y., Chan, C. K. Y., Good, D., Lee, B. P. Y., & Lai, V. K. W. (2018). Aspiring to become an engineer in Hong Kong: effects of engineering education and demographic background on secondary students' expectation to become an engineer. *European Journal of Engineering Education*, *43*(6), 824–841. https://doi.org/10.1080/03043797.2018.1435629

Lafuente-Ruiz-de-Sabando, A., Zorrilla, P., & Forcada, J. (2017). A review of higher education image and reputation literature: Knowledge gaps and a research agenda. *European Research on Management and Business Economics*. https://doi.org/10.1016/j.iedeen.2017.06.005

Lafuente-Ruiz-de-Sabando, A., Zorrilla, P., & Forcada, J. (2018). A review of higher education image and reputation literature: Knowledge gaps and a research agenda. *European Research on Management and Business Economics*, *24*(1), 8–16.

Lau, M. M. Y. (2016). Effects of 8Ps of services marketing on student selection of self-financing sub-degree programmes in Hong Kong. *International Journal of Educational Management*.

Le, T. D., Dobele, A. R., & Robinson, L. J. (2019). Information sought by prospective students from social media electronic word-of-mouth during the university choice process. *Journal of Higher Education Policy and Management*. https://doi.org/10.1080/1360080X.2018.1538595

Le, T. D., Robinson, L. J., & Dobele, A. R. (2020). Understanding high school students use of choice factors and word-of-mouth information sources in university selection. *Studies in Higher Education*, *45*(4), 808–818. https://doi.org/10.1080/03075079.2018.1564259

Lichtenberger, E., & George-Jackson, C. (2013a). Predicting High School Students' Interest in Majoring in a STEM Field: Insight into High School Students' Postsecondary Plans. *Journal of Career and Technical Education*, *28*(1), 19–38.

Lichtenberger, E., & George-Jackson, C. (2013b). Predicting High School Students' Interest in Majoring in a STEM Field: Insight into High School Students' Postsecondary Plans. *Journal of Career and Technical Education*, *28*(1), 19–38. http://proxy.geneseo.edu:2048/login?url=http://search.ebscohost.com/login.aspx?direct=true&db=eric&AN=EJ1043177&site=ehost-live&scope=site

Liguori, E., & Winkler, C. (2020). From Offline to Online: Challenges and Opportunities for Entrepreneurship Education Following the COVID-19 Pandemic. *Entrepreneurship Education and Pedagogy*, *3*(4), 346–351. https://doi.org/10.1177/2515127420916738

Lim, L. L., & Zailani, S. H. M. (2012). Determinants influencing intention to enrol on an online MBA programme. *International Journal of Business Information Systems*, *9*(1), 51–88. https://doi.org/10.1504/IJBIS.2012.044455

López Turley, R. N. (2009). College proximity: Mapping access to opportunity. *Sociology of Education*, *82*(2), 126–146. https://doi.org/10.1177/003804070908200202

Loyalka, P., Carnoy, M., Froumin, I., Dossani, R., Tilak, J. B., & Yang, P. (2014). Factors affecting the quality of engineering education in the four largest emerging economies. *Higher Education*, *68*(6), 977–1004. https://doi.org/10.1007/s10734-014-9755-8

Macleod, J., Kyriakidis, S., Kefallonitis, E., & Kavoura, A. (2017). Strategic innovative communication tools in higher education. In *Strategic Innovative Marketing* (pp. 759–764). Springer.

MacLeod, W. B., Riehl, E., Saavedra, J. E., & Urquiola, M. (2017). The big sort: College reputation and labor market outcomes. *American Economic Journal: Applied Economics*, *9*(3), 223–261. https://doi.org/10.1257/app.20160126

Mae, S. (2020). *Higher Ambitions: How American Plans for Post-Secondary Education*.

Magnell, M., Geschwind, L., & Kolmos, A. (2017). Faculty perspectives on the inclusion of work-related learning in engineering curricula. *European Journal of Engineering Education*, *42*(6), 1038–1047. https://doi.org/10.1080/03043797.2016.1250067

Mahajan, R., Agrawal, R., Sharma, V., & Nangia, V. (2014a). Factors affecting quality of management education in India: An interpretive structural modelling approach. *International Journal of Educational Management*, *28*(4), 379–399. https://doi.org/10.1108/IJEM-10-2012-0115

Mahajan, R., Agrawal, R., Sharma, V., & Nangia, V. (2014b). Factors affecting quality of management education in India. *International Journal of Educational Management*, *28*(4), 379–399. https://doi.org/10.1108/IJEM-10-2012-0115

Malgwi, C. A., Howe, M. A., & Burnaby, P. A. (2005a). Influences on students' choice of college major. *Journal of Education for Business*, *80*(5), 275–282.

Malgwi, C. A., Howe, M. A., & Burnaby, P. A. (2005b). Influences on Students' Choice of College Major. *Journal of Education for Business*, *80*(5), 275–282. https://doi.org/10.3200/JOEB.80.5.275-282

Mandelbaum, D. G. (1970). Society in India. Berkeley. *MandelbaumSociety in India1970*.

Mansfield, P. M., & Warwick, J. (2006). Gender differences in students' and parents' evaluative criteria when selecting a college. *Journal of Marketing for Higher Education*, *15*(2), 47–80.

Manzoor, S. R., Ho, J. S. Y., & Al Mahmud, A. (2020). Revisiting the 'university image model' for higher education institutions' sustainability. *Journal of Marketing for Higher Education*, *0*(0), 1–20. https://doi.org/10.1080/08841241.2020.1781736

María Cubillo-Pinilla, J., Sánchez-Herrera, J., & Pérez-Aguir, W. S. (2006). The influence of aspirations on higher education choice: a telecommunication engineering perspective. *European Journal of Engineering Education*, *31*(4), 459–469. https://doi.org/10.1080/03043790600676570

Maringe, F. (2006a). University and course choice: Implications for positioning, recruitment and marketing. *International Journal of Educational Management*, *20*(6), 466–479. https://doi.org/10.1108/09513540610683711

Maringe, F. (2006b). University and course choice. *International Journal of Educational Management*, *20*(6), 466–479. https://doi.org/10.1108/09513540610683711

Maringe, F., & Gibbs, P. (1989). Marketing Higher Education. In *McGraw-Hill House*. https://doi.org/10.1111/j.1468-2273.1989.tb01499.x

Maringe, F., & Gibbs, P. (2009). *Marketing Higher Education: Theory and Practice*. McGraw-Hill House. http://www.theeuropeanlibrary.org/tel4/record/2000007404839

Marinoni, G., Van't Land, H., & Jensen, T. (2020). The impact of Covid-19 on higher education around the world. *IAU Global Survey Report*.

Markes, I. (2006). A review of literature on employability skill needs in engineering. *European Journal of Engineering Education*, *31*(6), 637–650. https://doi.org/10.1080/03043790600911704

Martin, L., & Dixon, C. (2016). Making as a pathway to engineering. *Makeology: Makers as Learners*, *2*, 183.

Matusovich, H., Gillen, A., Carrico, C., Knight, D., & Grohs, J. (2020a). Outcome expectations and environmental factors associated with engineering college-going: A case study. *Journal of Pre-College Engineering Education Research (J-PEER)*, *10*(1).

Matusovich, H., Gillen, A., Carrico, C., Knight, D., & Grohs, J. (2020b). Outcome expectations and environmental factors associated with engineering college-going: A case study. *Journal of Pre-College Engineering Education Research*, *10*(1). https://doi.org/10.7771/2157-9288.1236

Maxwell Scott, E. (2000). Sample Size and Multiple Regression Analysis. *Psychological Methods*, *5*(4), 434–458.

Mazzarol, T., & Soutar, G. N. (2002). "Push-pull" factors influencing international student destination choice. *International Journal of Educational Management*, *16*(2), 82–90. https://doi.org/10.1108/09513540210418403

McCarthy, E. E., Sen, A. K., & Fox Garrity, B. (2012). Factors that influence Canadian students' choice of higher education institutions in the United States. *Business Education & Accreditation*, *4*(2), 85–95.

McDonough, P. M. (1997). *Choosing colleges: How social class and schools structure opportunity*. Suny Press.

McInerney, D. M. (2013). *Educational psychology: Constructing learning*. Pearson Higher Education AU.

Micari, M., & Pazos, P. (2012). Connecting to the Professor: Impact of the Student–Faculty Relationship in a Highly Challenging Course. *College Teaching*, *60*(2), 41–47. https://doi.org/10.1080/87567555.2011.627576

Miles, J. (2014). Tolerance and variance inflation factor. *Wiley StatsRef: Statistics Reference Online*.

Ming, J. S. K. (2010). Institutional factors influencing students' college choice decision in Malaysia: A conceptual framework. *International Journal of Business and Social Science*, *1*(3).

Mishkin, H., Wangrowicz, N., Dori, D., & Dori, Y. J. (2016). Career Choice of

Undergraduate Engineering Students. *Procedia - Social and Behavioral Sciences*, *228*(June), 222–228. https://doi.org/10.1016/j.sbspro.2016.07.033

Mitchell, J. E., Nyamapfene, A., Roach, K., & Tilley, E. (2019). Faculty wide curriculum reform: the integrated engineering programme. *European Journal of Engineering Education*, *0*(0), 1–19. https://doi.org/10.1080/03043797.2019.1593324

Moakler, M. W., & Kim, M. M. (2014). College major choice in STEM: Revisiting confidence and demographic factors. *Career Development Quarterly*, *62*(2), 128–142. https://doi.org/10.1002/j.2161-0045.2014.00075.x

Mogaji, E., & Maringe, F. (2020). Higher Education Marketing in Africa. In *Higher Education Marketing in Africa*. https://doi.org/10.1007/978-3-030-39379-3

Mok, K. H., Xiong, W., Ke, G., & Cheung, J. O. W. (2021). Impact of COVID-19 pandemic on international higher education and student mobility: Student perspectives from mainland China and Hong Kong. *International Journal of Educational Research*, *105*, 101718.

Moogan, Y. J., & Baron, S. (2003). An analysis of student characteristics within the student decision making process. *Journal of Further and Higher Education*, *27*(3), 271–287. https://doi.org/10.1080/0309877032000098699

Mourad, M. (2011). Role of brand related factors in influencing students' choice in Higher Education (HE) market. *International Journal of Management in Education*, *5*(2/3), 258. https://doi.org/10.1504/IJMIE.2011.039488

Munisamy, S., Mohd Jaafar, N. I., & Nagaraj, S. (2014). Does Reputation Matter? Case Study of Undergraduate Choice at a Premier University. *Asia-Pacific Education Researcher*, *23*(3), 451–462. https://doi.org/10.1007/s40299-013-0120-y

Murphy, G., & Salomone, S. (2013). Using social media to facilitate knowledge transfer in complex engineering environments: A primer for educators. *European Journal of Engineering Education*, *38*(1), 70–84. https://doi.org/10.1080/03043797.2012.742871

Murphy, M. P. A. (2020). COVID-19 and emergency eLearning: Consequences of the securitization of higher education for post-pandemic pedagogy. *Contemporary Security Policy*, *41*(3), 492–505. https://doi.org/10.1080/13523260.2020.1761749

Mwangi, C. A. G., Cabrera, A. F., & Kurban, E. R. (2019). Connecting school and home: Examining parental and school involvement in readiness for college through multilevel SEM. *Research in Higher Education*, *60*(4), 553–575.

Nandy, M., Lodh, S., & Tang, A. (2021). Lessons from Covid-19 and a resilience model for higher education. *Industry and Higher Education*, *35*(1), 3–9. https://doi.org/10.1177/0950422220962696

Natarajan, R. (2009). Assessment of engineering education quality: an Indian perspective. In *Engineering Education Quality Assurance* (pp. 145–151). Springer.

Neuman, W. L. (2013). *Social Research Methods: Qualitative and Quantitative Approaches* (Pearson education. (ed.); Pearson ed). Pearson education.

Neuwirth, L. S., Jović, S., & Mukherji, B. R. (2020). Reimagining higher education during and post-COVID-19: Challenges and opportunities. *Journal of Adult and Continuing Education*, *2059*. https://doi.org/10.1177/1477971420947738

Nieuwsma, J. A., Pepper, C. M., Maack, D. J., & Birgenheir, D. G. (2011). Indigenous perspectives on depression in rural regions of India and the United States. *Transcultural Psychiatry*, *48*(5), 539–568. https://doi.org/10.1177/1363461511419274

Nora, A., & Cabrera, A. F. (1992). Measuring Program Outcomes: What Impacts are Important to Assess and What Impacts are Possible to Nleasure? *PUB DATE 93 CONTRACT LC89082001 NOTE 172p. PUB TYPE Collected Works Conference Procee4ings (021)*, 86.

Nunnally, J. C., & Bernstein, I. H. (1967). *Psychometric theory* (Vol. 226). McGraw-Hill New York.

Nyaribo, M., Prakash, A., & Edward, O. (2012). Motivators of choosing a management course: A comparative study of kenya and india. *International Journal of Management Education*, *10*(3), 201–214. https://doi.org/10.1016/j.ijme.2012.08.001

Obermeit, K. (2012). Students' choice of universities in Germany: structure, factors and information sources used. *Journal of Marketing for Higher Education*, *22*(2), 206–230. https://doi.org/10.1080/08841241.2012.737870

Owlia, M. S., & Aspinwall, E. M. (1998). A framework for measuring quality in engineering education. *Total Quality Management*, *9*(6), 501–518. https://doi.org/10.1080/0954412988433

Pallant, J. (2005). *SPSS SURVIVAL MANUAL: A step by step guide to data analysis using SPSS for Windows (Version 12)*.

Pallant, Julie. (2020). *SPSS survival manual: A step by step guide to data analysis using IBM SPSS*. Routledge.

Palmer, A. (2003). The marketing of services. *The Marketing Book*, 585.

Palmer, A., Koenig-Lewis, N., & Asaad, Y. (2016). Brand identification in higher education: A conditional process analysis. *Journal of Business Research*, *69*(8), 3033–3040. https://doi.org/10.1016/j.jbusres.2016.01.018

Palmer, M., Hayek, J., Hossler, D., Jacob, S. A., Cummings, H., & Kinzie, J. (2004). *Fifty years of college choice: Social, political and institutional influences on the decision-making process*.

Pandi, A. P., Sethupathi, P. V. R., & Jeyathilagar, D. (2014). IEQMS model: a leveraging mechanism to polarise quality in engineering educational institutions - an empirical study. *International Journal of Manufacturing Technology and Management*, *28*(4/5/6), 257. https://doi.org/10.1504/IJMTM.2014.066703

Paulsen, M. B. (1990a). *College Choice: Understanding Student Enrollment Behavior*. ERIC.

Paulsen, M. B. (1990b). College choice: understanding student enrollment behavior. In *ASHE-ERIC higher education report CN - LB2350.5 .P38 1990* (Issue 1). School of Education and Human Development, George Washington University. http://files.eric.ed.gov/fulltext/ED333855.pdf

Paulsen, M. B. (1990c). College choice: understanding student enrollment behavior. In *ASHE-ERIC higher education report CN - LB2350.5 .P38 1990* (Issue 1). http://files.eric.ed.gov/fulltext/ED333855.pdf

Peters, J. (2018). *Designing Inclusion Into Engineering Education: A Fresh, Practical Look at how Diversity Impacts on Engineering and Strategies for Change*. Royal Academy of Engineering.

Phelps, L. A., Camburn, E. M., & Min, S. (2018). Choosing stem college majors: Exploring the role of pre-college engineering courses. *Journal of Pre-College Engineering Education Research*, *8*(1), 1–24. https://doi.org/10.7771/2157-9288.1146

Plewa, C., Ho, J., Conduit, J., & Karpen, I. O. (2016). Reputation in higher education: A fuzzy set analysis of resource configurations. *Journal of Business Research*. https://doi.org/10.1016/j.jbusres.2016.01.024

Porter, S. R., & Umbach, P. D. (2006). College major choice: An analysis of person-environment fit. *Research in Higher Education*, *47*(4), 429–449. https://doi.org/10.1007/s11162-005-9002-3

Powell, A., Dainty, A., & Bagilhole, B. (2012). Gender stereotypes among women engineering and technology students in the UK: Lessons from career choice narratives. *European Journal of Engineering Education*, *37*(6), 541–556. https://doi.org/10.1080/03043797.2012.724052

Powers, D. A., & Xie, Y. (1999). Statistical methods for categorical data analysis. *Statewide Agricultural Land Use Baseline 2015*, *1*, 309. https://doi.org/10.1017/CBO9781107415324.004

Price, I. F., Matzdorf, F., Smith, L., & Agahi, H. (2003). The impact of facilities on student choice of university. *Facilities*.

Price, I., Matzdorf, F., Smith, L., & Agahi, H. (2003). The impact of facilities on student choice of university. *Facilities*, *21*(10), 212–222. https://doi.org/10.1108/02632770310493580

Pucciarelli, F., & Kaplan, A. (2016). Competition and strategy in higher education: Managing complexity and uncertainty. *Business Horizons*, *59*(3), 311–320. https://doi.org/10.1016/j.bushor.2016.01.003

Raaper, R., & Brown, C. (2020). The Covid-19 pandemic and the dissolution of the university campus: implications for student support practice. *Journal of Professional Capital and Community*, *5*(3–4), 343–349. https://doi.org/10.1108/JPCC-06-2020-0032

Rajasenan, D. (2014). Gender Bias and Caste Exclusion in Engineering Admission. *Higher Education for the Future*, *1*(1), 11–28. https://doi.org/10.1177/2347631113518275

Rana, S., Anand, A., Prashar, S., & Haque, M. M. (2020). A perspective on the positioning of Indian business schools post COVID-19 pandemic. *International Journal of Emerging Markets*. https://doi.org/10.1108/IJOEM-04-2020-0415

Rashid, S., & Yadav, S. S. (2020). Impact of Covid-19 Pandemic on Higher Education and Research. *Indian Journal of Human Development*, *14*(2), 340–343. https://doi.org/10.1177/0973703020946700

Reay, D., Davies, J., David, M., & Ball, S. J. (2001). Choices of degree or degrees of choice? Class,'race'and the higher education choice process. *Sociology*, *35*(4), 855–874.

Regan, E., & DeWitt, J. (2015a). Attitudes, interest and factors influencing STEM enrolment behaviour: An overview of relevant literature. In *Understanding student participation*

*and choice in science and technology education* (pp. 63–88). Springer. https://doi.org/10.1007/978-94-007-7793-4_5

Regan, E., & DeWitt, J. (2015b). Attitudes, interest and factors influencing STEM enrolment behaviour: An overview of relevant literature. In *Understanding student participation and choice in science and technology education* (pp. 63–88). Springer.

Rojewski, J. (2002). Preparing the workforce of tomorrow: A conceptual framework for career and technical education. *Journal of Vocational Education Research*, *27*(1), 7–35.

Rosenberg, H., Nath, A., Leppard, J., & Syed, S. (2020). New challenges and mitigation strategies for resident selection during the coronavirus disease pandemic. *Canadian Journal of Emergency Medicine*, 1–3.

Rosenthal, G. T., Boudreaux, M., Boudreaux, D. L., Soignier, R. D., Folse, E., Frias, T., & Soper, B. (2014). The Student Storm Survey©: College Students' Thoughts on Their University's Response to a Natural Disaster. *Journal of Academic Administration in Higher Education*, *10*(2), 19–26.

Rutter, R., Roper, S., & Lettice, F. (2016). Social media interaction, the university brand and recruitment performance. *Journal of Business Research*. https://doi.org/10.1016/j.jbusres.2016.01.025

Sahney, S., Banwet, D. K., & Karunes, S. (2004). A SERVQUAL and QFD approach to total quality education. *International Journal of Productivity and Performance Management*, *53*(2), 143–166. https://doi.org/10.1108/17410400410515043

Sahu, A. R., Shrivastava, R. R., & Shrivastava, R. L. (2013). Critical success factors for sustainable improvement in technical education excellence. *The TQM Journal*, *25*(1), 62–74. https://doi.org/10.1108/17542731311286432

Sakthivel, P. B., & Raju, R. (2006a). *An Instrument for Measuring Engineering Education Quality from Students ' Perspective*. *6967*. https://doi.org/10.1080/10686967.2006.11918559

Sakthivel, P. B., & Raju, R. (2006b). Conceptualizing total quality management in engineering education and developing a TQM educational excellence model. *Total Quality Management and Business Excellence*, *17*(7), 913–934. https://doi.org/10.1080/14783360600595476

Salami, S. O. (2007a). Influence of culture, family and individual differences on choice of gender-dominated occupations among female students in tertiary institutions. *Women in Management Review*, *22*(8), 650–665. https://doi.org/10.1108/09649420710836326

Salami, S. O. (2007b). Influence of culture, family and individual differences on choice of gender-dominated occupations among female students in tertiary institutions. *Women in Management Review*.

Sánchez-Gordón, M., & Colomo-Palacios, R. (2020). Factors influencing software engineering career choice of Andean indigenous. *Proceedings of the ACM/IEEE 42nd International Conference on Software Engineering: Companion Proceedings*, 264–265.

Sanders, J. (2005). Gender and technology in education: A research review. *Seattle: Center for Gender Equity. Bibliography Retrieved March*, *20*, 2006.

Saranraj, L., Khan, Z. A., & Zafar, S. (2016). Influence of motivational factors and gender

differences on learning english as a second language: A case of engineering students from rural background. *Indian Journal of Science and Technology*, *9*(44). https://doi.org/10.17485/ijst/2016/v9i44/99721

Saunders-Smits, G., & de Graaff, E. (2012). Assessment of curriculum quality through alumni research. *European Journal of Engineering Education*, *37*(2), 133–142. https://doi.org/10.1080/03043797.2012.665847

Sayeda, B., Rajendran, C., & Sai Lokachari, P. (2010). An empirical study of total quality management in engineering educational institutions of India. *Benchmarking: An International Journal*, *17*(5), 728–767. https://doi.org/10.1108/14635771011076461

Schuh, J. H., & Santos Laanan, F. (2006). Forced transitions: The impact of natural disasters and other events on college students. *New Directions for Student Services*, *2006*(114), 93–102.

Settersten, R. A., Bernardi, L., Härkönen, J., Antonucci, T. C., Dykstra, P. A., Heckhausen, J., Kuh, D., Mayer, K. U., Moen, P., Mortimer, J. T., Mulder, C. H., Smeeding, T. M., van der Lippe, T., Hagestad, G. O., Kohli, M., Levy, R., Schoon, I., & Thomson, E. (2020). Understanding the effects of Covid-19 through a life course lens. *Advances in Life Course Research*, *45*(July), 100360. https://doi.org/10.1016/j.alcr.2020.100360

Shay, S. (2014). Curriculum in higher education: Beyond false choices. In *Thinking about higher education* (pp. 139–155). Springer.

Sia, J. K.-M., & Adamu, A. A. (2020). Facing the unknown: pandemic and higher education in Malaysia. *Asian Education and Development Studies*.

Sia, J. K. M., & Ming, K. E. E. (2010). *A model of higher education institutions choice in Malaysia-A conceptual approach*. College of Law, Government and International Studies, Universiti Utara Malaysia.

Simões, C., & Soares, A. M. (2010). Applying to higher education: Information sources and choice factors. *Studies in Higher Education*, *35*(4), 371–389. https://doi.org/10.1080/03075070903096490

Singer, T. S., & Hughey, A. W. (2002). The role of the alumni association in student life. *New Directions for Student Services*, *2002*(100), 51–68. https://doi.org/10.1002/ss.70

Singh, K., Allen, K. R., Scheckler, R., & Darlington, L. (2007). Women in computer-related majors: A critical synthesis of research and theory from 1994 to 2005. *Review of Educational Research*, *77*(4), 500–533. https://doi.org/10.3102/0034654307309919

Sohrabi, C., Alsafi, Z., O'Neill, N., Khan, M., Kerwan, A., Al-Jabir, A., Iosifidis, C., & Agha, R. (2020). World Health Organization declares global emergency: A review of the 2019 novel coronavirus (COVID-19). *International Journal of Surgery*, *76*, 71–76.

Sojkin, B., Bartkowiak, P., & Skuza, A. (2012). Determinants of higher education choices and student satisfaction: The case of Poland. *Higher Education*, *63*(5), 565–581. https://doi.org/10.1007/s10734-011-9459-2

Sood, S., & Sharma, A. (2021). *Resilience and Psychological Well-Being of Higher Education Students During COVID-19 : The Mediating Role of Perceived Distress*. *2020*(March 2020). https://doi.org/10.1177/0972063420983111

Soutar, G. N., & Turner, J. P. (2002). Students' preferences for university: a conjoint

analysis. *International Journal of Educational Management*, *16*(1), 40–45. https://doi.org/10.1108/09513540210415523

Sovansophal, K. (2019). Family socioeconomic status and students' choice of STEM majors: Evidence from higher education of Cambodia. *International Journal of Comparative Education and Development*, *22*(1), 49–65. https://doi.org/10.1108/IJCED-03-2019-0025

St. John, E. P., Paulsen, M. B., & Carter, D. F. (2005). Diversity, college costs, and postsecondary opportunity: An examination of the financial nexus between college choice and persistence for African Americans and Whites. *Journal of Higher Education*, *76*(5), 545–569. https://doi.org/10.1080/00221546.2005.11772298

Stange, K. (2015). Differential pricing in undergraduate education: Effects on degree production by field. *Journal of Policy Analysis and Management*, *34*(1), 107–135.

Steckler, A., McLeroy, K. R., Goodman, R. M., Bird, S. T., & McCormick, L. (1992). Toward integrating qualitative and quantitative methods: An introduction. *Health Educ Q*, *19*, 1–8. https://doi.org/10.1177/109019819201900101

Stevens, J. P. (2012). *Applied multivariate statistics for the social sciences*. Routledge.

Sulcic, V. (2010). The key factors for acquired knowledge through e-learning. *International Journal of Innovation and Learning*, *7*(3), 290–302.

Sun, L. B., & Qu, H. (2011). Is there any gender effect on the relationship between service quality and word-of-mouth? *Journal of Travel and Tourism Marketing*, *28*(2), 210–224. https://doi.org/10.1080/10548408.2011.546215

Swan, A. (2015). Experiential and contextual factors that shape engineering interest and educational decision-making processes among female students. *NASPA Journal About Women in Higher Education*, *8*(1), 82–100.

Tan, A. (2020). *Zoom is your new classroom: Will online education become the norm after COVID-19?*

Tarafdar, M., & Zhang, J. (2005). Analysis of critical website characteristics: a cross-category study of successful websites. *Journal of Computer Information Systems*, *46*(2), 14–24. https://doi.org/10.1080/08874417.2006.11645879

Tarc, P. (2020). Education post-'Covid-19': Re-visioning the face-to-face classroom. *Current Issues in Comparative Education (CICE)*, *22*(1).

Tas, A., & Ergin, E. A. (2012). Key factors for student recruitment: The issue of university branding. *International Business Research*, *5*(10), 146.

Telli Yamamoto, G. (2006). University evaluation-selection: a Turkish case. *International Journal of Educational Management*, *20*(7), 559–569. https://doi.org/10.1108/09513540610704654

Tesar, M. (2020). Towards a Post-Covid-19 'New Normality?': Physical and Social Distancing, the Move to Online and Higher Education. *Policy Futures in Education*, *18*(5), 556–559. https://doi.org/10.1177/1478210320935671

Thompson, R. (2009). Social class and participation in further education: evidence from the Youth Cohort Study of England and Wales. *British Journal of Sociology of Education*,

*30*(1), 29–42.

Tight, M. (2015). *Researching Higher Education*. https://doi.org/10.4324/9781315675404

Tilak, J. B. G. (2020). Determinants of Students' Choice of Engineering Disciplines in India. *Yuksekogretim Dergisi*, *10*(2), 163–180. https://doi.org/10.2399/yod.19.017000

Toquero, C. M. (2020). Challenges and Opportunities for Higher Education amid the COVID-19 Pandemic: The Philippine Context. *Pedagogical Research*, *5*(4), em0063. https://doi.org/10.29333/pr/7947

Toutkoushian, R., & Paulsen, M. (2016). Economics of Higher Education. In *Economics of Higher Education: Background, Concepts, and Applications*. https://doi.org/10.1007/978-94-017-7506-9

Trum, H. (1992). Aspects of quality in Continuing Engineering Education. *International Journal of Continuing Engineering Education and Life-Long Learning*, *2*(1), 1–13. https://www.scopus.com/inward/record.uri?eid=2-s2.0-0026706012&partnerID=40&md5=126ad928a31ed911b4aad32650fdde2d

Turley, R. N. L. (2009). College proximity: Mapping access to opportunity. *Sociology of Education*, *82*(2), 126–146.

Turner, M. L. (2017). Like, Love, Delete: Social Media's Influence on College Choice. *Journal of College Admission*, *237*, 31–33.

Turner, S. E., & Bowen, W. G. (1999). Choice of major: The changing (unchanging) gender gap. *ILR Review*, *52*(2), 289–313.

UGC. (2021a). *UGC guidelines for re-opening the universities and colleges post lockdown due to Covid-19 pandemic*.

UGC. (2021b). *UGC guidelines for re-opening the universities and colleges post lockdown due to Covid-19 pandemic*. https://www.ugc.ac.in/ugc_notices.aspx

UNESCO. (2010). Engineering: Issues Challenges and Opportunities for Development. In *Engineering : Issues Challenges and Opportunities for Development*. http://unesdoc.unesco.org/images/0018/001897/189753e.pdf

UNESCO. (2020). Covid-19 crisis and curriculum: sustinaining quality outcomes in the context of remote learning. *UNESCO Covid-19 Eduaction Response. Education Sector Issue Notes. Issue Note No. 4.2*, *4.2*, 1–6. https://unesdoc.unesco.org/ark:/48223/pf0000373273

Upadhayay, L., & Vrat, P. (2017). Quality issues affecting organisational performance : a review of Indian technical education. *Int. J. Indian Culture and Business Management*, *14*(1).

Van Teijlingen, E. R., & Hundley, V. (2001). *The importance of pilot studies*.

Veloutsou, C., Paton, R. A., & Lewis, J. (2005a). Consultation and reliability of information sources pertaining to university selection: Some questions answered? *International Journal of Educational Management*, *19*(4), 279–291. https://doi.org/10.1108/09513540510599617

Veloutsou, C., Paton, R. A., & Lewis, J. (2005b). Consultation and reliability of information sources pertaining to university selection. *International Journal of Educational*

*Management*.

Viswanadhan, K. G. (2009). Quality indicators of engineering education programmes: a multi-criteria analysis from India. *International Journal of Industrial and Systems Engineering*, *4*(3), 270–282.

Voss, R., Gruber, T., & Szmigin, I. (2007). Service quality in higher education: The role of student expectations. *Journal of Business Research*, *60*(9), 949–959. https://doi.org/10.1016/j.jbusres.2007.01.020

Wadhwa, R. (2016a). Students on move: Understanding decision-making process and destination choice of Indian students. *Higher Education for the Future*, *3*(1), 54–75.

Wadhwa, R. (2016b). Students on Move: Understanding Decision-making Process and Destination Choice of Indian Students. *Higher Education for the Future*, *3*(1), 54–75. https://doi.org/10.1177/2347631115610221

Walkington, J. (2002). A process for curriculum change in engineering education. *European Journal of Engineering Education*, *27*(2), 133–148. https://doi.org/10.1080/03043790210129603

Weisser, R. A. (2020). How Personality Shapes Study Location Choices. *Research in Higher Education*, *61*(1), 88–116. https://doi.org/10.1007/s11162-019-09550-2

Whiston, S. C., & Keller, B. K. (2004). The Influences of the Family of Origin on Career Development: A Review and Analysis. *The Counseling Psychologist*, *32*(4), 493–568. https://doi.org/10.1177/0011000004265660

WHO. (2020). *Checklist to support schools re-opening and preparation for COVID-19 resurgences or similar public health crises*.

Wladis, C., Wladis, K., & Hachey, A. C. (2014a). The Role of Enrollment Choice in Online Education: Course Selection Rationale and Course Difficulty as Factors Affecting Retention. *Online Learning*, *18*(3), n3.

Wladis, C., Wladis, K., & Hachey, A. C. (2014b). The role of enrollment choice in online education: Course selection rationale and course difficulty as factors affecting retention. *Journal of Asynchronous Learning Network*, *18*(3), 1–14. https://doi.org/10.24059/olj.v18i3.391

Wong, P., Ng, P. M. L., Lee, D., & Lam, R. (2019). Examining the impact of perceived source credibility on attitudes and intentions towards taking advice from others on university choice. *International Journal of Educational Management*, *34*(4), 709–724. https://doi.org/10.1108/IJEM-06-2019-0190

Woolnough, B. E. (1994). Factors affecting students' choice of science and engineering. *International Journal of Science Education*, *16*(6), 659–679. https://doi.org/10.1080/0950069940160605

Wu, F., Fan, W., Arbona, C., & de la Rosa-Pohl, D. (2020). Self-efficacy and subjective task values in relation to choice, effort, persistence, and continuation in engineering: an Expectancy-value theory perspective. *European Journal of Engineering Education*, *45*(1), 151–163. https://doi.org/10.1080/03043797.2019.1659231

Xiong, W., MOK, K. H. J., Guoguo, K. E., & CHEUNG, O. W. J. (2020). *Impact of COVID-19 Pandemic on International Higher Education and Student Mobility: Student*

*Perspectives from Mainland China and Hong Kong*.

Yamamoto, G. T. (2006a). University evaluation-selection: A Turkish case. *International Journal of Educational Management*, *20*(7), 559–569. https://doi.org/10.1108/09513540610704654

Yamamoto, G. T. (2006b). University evaluation-selection: a Turkish case. *International Journal of Educational Management*.

Yelkur, R. (2000). Customer Satisfaction and the Services Marketing Mix. *Journal of Professional Services Marketing*, *21*(1), 105–115. https://doi.org/10.1300/J090v21n01

Yoon, H. (2019). An online college near me: Exploring the institutional factors of e-learners' local orientation. *International Review of Research in Open and Distance Learning*, *20*(5), 64–84. https://doi.org/10.19173/irrodl.v20i5.4432

Young, D. J., Fraser, B. J., & Woolnough, B. E. (1997). Factors affecting student career choice in science: An Australian study of rural and urban schools. *Research in Science Education*, *27*(2), 195–214. https://doi.org/10.1007/BF02461316

Yusuf, B. N., & Jihan, A. (2020). Are we prepared enough? A case study of challenges in online learning in a private higher learning institution during the Covid-19 outbreaks. *Advances in Social Sciences Research Journal*, *7*(5), 205–212.

Zia, A. (2020a). Exploring factors influencing online classes due to social distancing in COVID-19 pandemic: a business students perspective. *International Journal of Information and Learning Technology*, *37*(4), 197–211. https://doi.org/10.1108/IJILT-05-2020-0089

Zia, A. (2020b). Exploring factors influencing online classes due to social distancing in COVID-19 pandemic: a business students perspective. *The International Journal of Information and Learning Technology*.

Zuhairi, A., Raymundo, M. R. D. R., & Mir, K. (2020). Implementing quality assurance system for open and distance learning in three Asian open universities: Philippines, Indonesia and Pakistan. *Asian Association of Open Universities Journal*, *15*(3), 297–320. https://doi.org/10.1108/aaouj-05-2020-0034